\documentclass[oldversion]{aa}

\usepackage{graphics}
\usepackage{txfonts}   
\usepackage[dvips]{epsfig}

\begin{document} 
 
   \title{The Hamburg/ESO R-process Enhanced Star survey (HERES) \newline
   \thanks{Based on observations carried out at the European Southern
           Observatory, Paranal, Chile. Proposal number (170.D-0010)},
   \thanks{Table 2 is only available in electronic form at the CDS via
           anonymous ftp to cdsarc.u-strasbg.fr (130.79.128.5) or via
           http://cdsweb.u-strasbg.fr/Abstract.html\,.}
         }
   \subtitle{III. HE~0338$-$3945 and the formation of the r+s stars}

   \author{K. Jonsell\inst{1}
   \and    P. S. Barklem\inst{1}
   \and    B. Gustafsson\inst{1}
   \and    N. Christlieb\inst{2}
   \and    V. Hill\inst{3}
   \and    T. C. Beers\inst{4}
   \and    J. Holmberg\inst{5}
}

   \offprints{K. Jonsell}

   \institute{Department of Astronomy and Space Physics, Uppsala Astronomical
              Observatory, Box 515, S-751 20 Uppsala, Sweden
              (firstname.lastname@astro.uu.se)
   \and       Hamburger Sternwarte, Gojenbergsweg 112, D-21029 Hamburg, Germany.
   \and       GEPI, Observatoire de Paris-Meudon, F-92125 Meudon Cedex, France
   \and       Department of Physics and Astronomy and JINA: Joint Institute 
              for Nuclear Astrophysics, Michigan State University, East Lansing, MI 48824 USA
   \and       Tuorla Observatory, V{\"a}is{\"a}l{\"a}ntie 20, FI-21500 Piikki{\"o}, Finland 
 }

   \date{Received 4 November 2005 / Accepted 3 January 2006}

   \authorrunning{K. Jonsell et al.}

   \abstract{
We have derived abundances of 33 elements and upper limits for 6 additional elements for the metal-poor ($\mathrm{[Fe/H]} = -2.42$) turn-off star HE~0338$-$3945 from high-quality VLT-UVES spectra. The star is heavily enriched, by about a factor of 100 relative to iron and the Sun, in the heavy $s$-elements (Ba, La, ..).  It is also heavily enriched in Eu, which is generally considered an $r$-element, and in other similar elements. It is less enriched, by about a factor of 10, in the lighter $s$-elements (Sr, Y and Zr). C is also strongly enhanced and, to a somewhat lesser degree, N and O.  These abundance estimates are subject to severe uncertainties due to NLTE and thermal inhomogeneities which are not taken into detailed consideration. However, an interesting result, which is most probably robust in spite of these uncertainties, emerges: the abundances derived for this star are very similar to those of other stars with an overall enhancement of all elements beyond the iron peak. 

We have defined criteria for this class of stars, r+s stars, and discuss nine different scenarios to explain their origin. None of these explanations is found to be entirely convincing. The most plausible hypotheses involve a binary system in which the primary component goes through its giant branch and asymptotic giant branch phases and produces CNO and $s$-elements which are dumped onto the observed star. Whether the $r$-element Eu is produced by supernovae before the star was formed (perhaps triggering the formation of a low-mass binary), by a companion as it explodes as a supernova (possibly triggered by mass transfer), or whether it is possibly produced in a high-neutron-density version of the $s$-process is still unclear. Several suggestions are made on how to clarify this situation.

 }

     \maketitle

\section{Introduction}

Elements with atomic numbers $\mathrm{Z} > 30$ are believed to be almost exclusively synthesized in neutron-capture ($n$-capture) reactions. In the most metal-poor stars the overall abundance of these elements varies from star to star, by more than a factor of 100 at a given metallicity (McWilliam et~al.~\cite{mcwilliam1995}, Ryan et~al.~\cite{ryan1996}). Also, the different abundance ratios vary, e.g. the Ba/Eu ratio tends to decline with decreasing [Fe/H]\footnote{We use the standard notations $\mathrm{A/B} = N_\mathrm{A}/N_\mathrm{B} $ and $\mathrm{[A/B]} = \log ( N_\mathrm{A} / N_\mathrm{B} )_\star - \log (N_\mathrm{A} / N_\mathrm{B})_\mathrm{\sun}$ where $N_\mathrm{X}$ are number densities.} (McWilliam~\cite{mcwilliam1998}, Burris et~al.~\cite{burris2000}). Eu is often thought to be synthesized in conditions of very high neutron fluxes (the {\it r-process}) while Ba is most easily made in conditions of much lower neutron fluxes (the {\it s-process}). The decline of Ba/Eu may show the relative dominance of $r$-process sites in the early history of the Galaxy. 

The first Extremely Metal Poor (EMP, Beers \& Christlieb~\cite{beers2005}) star found to be enriched in $n$-capture elements was the giant HD\,115444 with $\mathrm{[Fe/H]} \sim -3$ (Griffin et~al.~\cite{griffin1982}). Westin et~al.~(\cite{westin2000}) found marginal overabundances of Ba and other $s$-elements relative to iron in this star as compared with the Sun. This still makes the star rich in these elements relative to ``normal'' Population\,{\sc ii} stars of the same metallicity. However, it has a high abundance of europium, $\mathrm{[Eu/Fe]} = 0.85$, even compared to the Sun. 

The heavy elements of HD\,115444 show good agreement in relative abundances with scaled solar $r$-process abundances, as does the EMP giant CS\,22892-052 which is even more $r$-element rich (McWilliam et~al.~\cite{mcwilliam1995}, Sneden et~al.~\cite{sneden2003}). This agreement suggests that a similar process is responsible for generating the heavy $r$-elements, both for the most metal-poor stars and for the Sun. However, several studies have found a deviation from the scaled solar $r$-process abundance pattern for the light $r$-process elements (e.g.\ Sneden et~al.~\cite{sneden2000}, \cite{sneden2003}, Cowan et~al.~\cite{cowan2002}, Aoki et~al.~\cite{aoki2005}). Also, similar stars like CS\,31082-001 (Hill et~al.~\cite{hill2002}) and CS\,30306-132 (Honda et~al.~\cite{honda2004a}) seem to have had an ``actinide boost'', which has selectively affected the abundances of the most heavy $r$-elements. These variations in behaviour of the $r$-elements suggest that multiple astrophysical sites for the $r$-process may exist.

The discovery of CS\,22892-052 was a result of the compilation of metal-poor HK survey objects of Beers et~al.~(\cite{beers1992}). Similarly, the large programme of Cayrel et~al.~(\cite{cayrel2004}) at ESO, ``First Stars'', was founded on that survey. In the First Stars project, high-resolution studies of the very metal-poor giant CS\,31082-001 revealed its high Eu abundance. It was possible to identify a \ion{U}{ii} line in this star, and this opened up the possibility to use the U/Th ratio as a new chronometer. More recently, Honda et~al.~(\cite{honda2004a, honda2004b}) reported on two new Eu-rich stars, again taken from the HK survey, namely CS\,30306-132  with $\mathrm{[Eu/Fe]} = +0.85$ and CS\,22183-031 with $\mathrm{[Eu/Fe]} = +1.2$. According to the classification suggested by Beers \& Christlieb (\cite{beers2005}) the latter of these is an r-II star (having $\mathrm{[Eu/Fe]} > +1.0$ and $\mathrm{[Ba/Eu]} < 0.0$) while the former is classified as r-I ($+0.3 \le \mathrm{[Eu/Fe]} \le +1.0$ and $\mathrm{[Ba/Eu]} < 0.0$). Another interesting object which is very overabundant in $n$-capture elements, the carbon-enhanced metal-poor (CEMP) star with $s$-element enhancement CS\,31062-050, was recently studied by Johnson \& Bolte~(\cite{johnson2004}). Interestingly, they concluded that the abundance profile of the star, in spite of its high Eu abundance, suggests that the $r$-process was not responsible in this case. These indications will be further discussed in Sect.\,\ref{discussion}.

Although the contribution of the $s$-process to the gas in the very early Galaxy may in general be tiny, as one would expect due to the small fraction of iron to start the nucleosynthesis from, some stars with [Fe/H] as low as about $-3$ have been found to have clear $s$-element signatures (Johnson \& Bolte~\cite{johnson2002a}, Simmerer et~al.~\cite{simmerer2004}, Sivarani et~al.~\cite{sivarani2004}). The implications of the presence of these stars are still not fully understood. It was predicted by Gallino et~al.~(\cite{gallino1998}) (see also Goriely \& Mowlavi~\cite{goriely2000}), that the fraction of the heaviest $s$-elements should be enhanced in metal-poor environments with respect to the lighter $s$-elements, a result of the scarcity of seed nuclei with respect to neutrons. This prediction was qualitatively confirmed by Aoki et~al.~(\cite{aoki2000}) who made the first discovery of Pb in a $s$-element rich very metal-poor star, LP625-44. Somewhat more iron-rich ``lead stars'' were also discovered by Van Eck et~al.~(\cite{van_eck2001}) (see also Sivarani et~al.~\cite{sivarani2004}). We also note that the abundances of the ``lighter'' $n$-capture elements (Sr, Y, Zr) do not scale very well with those of the heavier elements (Ba etc, Z $\ge 56$). This may be due to the fact that the lighter $s$-elements (the so-called ``weak component'', see Prantzos et~al.~\cite{prantzos1990}) received important contributions from seed nuclei reacting with neutrons created in the $^{22}$Ne($\alpha$,n)$^{25}$Mg reaction in massive AGB stars, while the heavy elements were essentially contributed by neutrons from the $^{13}$C($\alpha$,n)$^{16}$O reaction in less massive stars. 

Another class of stars was discovered by Barbuy et~al.~(\cite{barbuy1997}) and Hill et~al.~(\cite{hill2000}). Two carbon-rich Population\,{\sc ii} giants, the CEMP stars CS\,22948-027 and CS\,29497-034 from the HK survey, were found not only to be rich in the commonly enhanced $s$-elements such as Sr, Y, Ba and La, but also the $r$-element Eu was significantly enhanced. Preston \& Sneden~(\cite{preston2001}) found another star of this class in the HK survey (CS\,22898-027), Aoki et al. (\cite{aoki2002}) added two more (CS\,29526-110 and CS\,31062-012) and Sivarani et~al.~(\cite{sivarani2004}) another (CS\,29497-030) (see also Ivans et~al.~\cite{ivans2005}). The stars of this type presently known are listed in Table\,\ref{Tstars}. We shall denote these stars {\it r+s stars}, and we postpone a detailed discussion of the definition of this class of stars to Sect.\,\ref{discussion}.

The origin of the abundance peculiarities of the r+s stars is not clear, and many scenarios have been proposed (see Sect.\,\ref{discussion}). However, this is not the only reason for studying them further. In fact, the astrophysical sites of the $r$-process are still unclear. The precise sites of the $s$-process among the most metal-poor stars are also still debatable, as well as the fraction of nuclei provided by neutrons from the $^{13}$C($\alpha$,n)$^{16}$O reaction as compared with the $^{22}$Ne($\alpha$,n)$^{25}$Mg neutron source. One might hope that a clarification of the origin of the r+s stars may shed some light on the general questions concerning the sites of the $r$- and $s$-processes. 

The Hamburg/ESO objective-prism survey for bright quasars (HES; Wisotzki et~al.~\cite{wisotzki2000}) has been used for a search for metal-poor stars (Christlieb~\cite{christlieb2003}). One of these was the r+s star HE~2148$-$1247, found in the Keck pilot programme on EMP stars from the HES (Cohen et~al.~\cite{cohen2003}). In a systematic approach to exploring the $n$-enhanced HES stars, the Hamburg/ESO R-process Enhanced Star survey (HERES) has been carried out (Christlieb et~al.~\cite{christlieb2004}, hereafter Paper\,{\sc i}). For several hundred candidate stars from HES, ``snapshot'' spectra at moderate resolving power ($R \sim 20\,000$) and low $S/N$ ($\sim 50$) were obtained with the main goal to find $r$-element enriched stars, as disclosed by strong Eu lines. These spectra were analysed with an automatic procedure based on MARCS model atmospheres and synthetic spectra (Barklem et al.~\cite{barklem2005}, Paper\,{\sc ii}). As a result of the HERES survey, a number of new detections of $n$-capture enriched EMP stars were made. In particular, 8 new r-II stars and 35 r-I stars were found.

One of the stars found by the HERES survey is HE~0338$-$3945. This object is a star located close to the main-sequence turnoff of an old halo population with an overall metallicity of $\mathrm{[Fe/H]} \sim -2.4$, and enriched in both Ba and Eu. We decided to obtain spectra of higher quality of this star and perform a detailed analysis which is presented here. In Sects.\,\ref{observations} and \ref{analysis} the observations and analysis respectively are described.  In Sect.\,\ref{resultsA} the abundance results are presented.  In Sect.\,\ref{comparisons} the abundances are compared with those of other stars.  This reveals HE~0338$-$3945 to be very similar to another r+s star, HE~2148$-$1247 (Cohen~et~al.~\cite{cohen2003}).  This result led us to survey the known r+s stars in the literature, and this survey, along with a comparison with HE~0338$-$3945 is also presented.  In Sect.\,\ref{discussion} the results for HE~0338$-$3945 are discussed, as is both the classification and formation of r+s stars in general.  Finally, in Sect.\,\ref{conclusions} the conclusions are presented. 

\section{Observations and data reduction}
\label{observations}

The basic data for the star HE~0338$-$3945 are presented in Table\,\ref{Tstellardata}.  It was observed in Service Mode at the VLT Unit Telescope 2 (Kueyen) equipped with the spectrograph UVES at the European Southern Observatory at Paranal, Chile, during the nights of 11 December and 23 to 25 December, 2002.

\begin{table}
\begin{center}
\caption{Basic data and parameters for the star HE~0338$-$3945. ($^1$) Burstein \& Heiles~(\cite{burstein1982}). ($^2$) Schlegel et~al.~(\cite{schlegel1998}).}
\label{Tstellardata} 
\begin{tabular}{lll}
\hline
\hline
Quantity & Value & Uncertainty  \\ 
\hline 
R.A. (J2000)           	& $03\mathrm{h} 39\mathrm{m} 54.9\mathrm{s}$ & $\pm 0.1\mathrm{s}$  \\  
Decl. (J2000)          	& $-39\degr 35\arcmin 44\arcsec  $           & $\pm 1\arcsec$  \\  
$V$                    	& $15.333$~mag                               & $\pm 0.007$~mag   \\  
$B-V$                  	& $0.420$~mag                                & $\pm 0.016$~mag   \\
$V-R$                  	& $0.235$~mag                                & $\pm 0.011$~mag   \\
$V-I$                  	& $0.546$~mag                                & $\pm 0.010$~mag   \\
$J$                    	& $14.403$~mag                               & $\pm 0.032$~mag  \\
$H$                    	& $14.114$~mag                               & $\pm 0.035$~mag  \\
$K$                    	& $14.084$~mag                               & $\pm 0.059$~mag  \\
$E(B-V)^1$             	& $0.000$~mag                                &  \\
$E(B-V)^2$             	& $0.013$~mag                                &  \\
$T_\mathrm{eff}$       	& $6160$\,K                                  & $\pm 100$\,K  \\  
$\mathrm{[Fe/H]}    $  	& $-2.42$\,dex                               & $\pm 0.11$\,dex  \\  
$\log g$               	& $4.13$\,dex                                & $\pm 0.33$\,dex  \\  
$\xi_\mathrm{t}$       	& $1.13$\,km/s                               & $\pm 0.22$\,km/s   \\  
$\mathrm{V}_\mathrm{r}$ & $177.9$\,km/s                              & $\pm 0.5$ km/s  \\  
Class.                  & turn-off star                              &  \\  
\hline
\end{tabular}
\end{center}
\end{table}
 
To cover as large a spectral range as possible, UVES was used in dichroic mode with two different settings: Dichroic~1 with central wavelengths of 3460\,{\AA} (3055--3850\,{\AA}) and 5800\,{\AA} (4795--5745{\AA} and 5855--6755\,{\AA}) in the blue and red arm of the spectrograph respectively, and Dichroic~2 with central wavelengths of 4370\,{\AA} (3760--4945\,{\AA}) and 8600\,{\AA} (6720--8425{\AA} and 8710--10510\,{\AA}). The exposure times were in total 6 hours for the Dichroic~1 setting and 9.5 hours for the Dichroic~2 setting, consisting of single exposure times of 30 to 75 minutes. The projected slit width was set to 1.2\arcsec, yielding a resolving power of $R=\lambda/\Delta\lambda$ equal to $30,000-40,000$. During the observations, the CCD binning was set to $1\times 1$, but after reduction the spectra were rebinned by a factor of two to increase the signal-to-noise ratio ($S/N$) per pixel. Since the pixel scale after rebinning is $0.44\mbox{\arcsec}/\mbox{pixel}$ and $0.32\mbox{\arcsec}/\mbox{pixel}$ in the blue and red arm, respectively, the spectra are appropriately sampled.

The optical photometry data ($BVRI$) for HE~0338-3945 were obtained on the night of 24 October 2002 by J. Holmberg using the Danish 1.5m telescope with DFOSC on La Silla.  Details of the reduction and analysis of these data are provided by Beers et al. (2005, in preparation).  These data were supplemented by near-IR $JHK$ from the 2MASS Point Source Catalog (Cutri et~al.~\cite{cutri2003}).

The reduction was made in the standard way by using the IDL-package REDUCE, Piskunov \& Valenti~(\cite{piskunov2002}).
There were some problems with discontinuities and other reduction artifacts, cosmic ray hits, and noise in the spectrum, so the usable spectral ranges were finally 3100--5741\,{\AA} and 5847--8487\,{\AA}, now given in the rest frame of the star. The $S/N$ in the spectra was rather poor at the blue end. At 3700\,{\AA} it had risen to about 40, and from 4600\,{\AA} it was 70 or higher, peaking at 6700\,{\AA} at 130. The $S/N$ was about 74 in mean for the final spectra.

\section{Abundance analysis}
\label{analysis}

\subsection{Fundamental parameters and atmospheric model}
\label{parametersmodel}

In Paper\,{\sc ii}, a temperature of $6162\pm 100$\,K was derived from photometry when adopting reddening derived from maps of Schlegel~et~al.~(\cite{schlegel1998}). A subsequent analysis of the snapshot spectrum gave $\log g = 4.09$ and $\mathrm{[Fe/H]} = -2.41$. We note that if instead the reddening maps of Burstein \& Heiles~(\cite{burstein1982}) were adopted the temperature would be 70\,K lower. 

The effective temperature for analysis of the higher quality spectrum was redetermined from H$\beta$ and H$\delta$.  The continuum rectification and analysis, including error estimation, were performed following Barklem et~al.~(\cite{barklem2002}).  These two lines were employed as the continuum rectification method was able to be successfully applied.  The gravity and metallicity from Paper\,{\sc ii} were adopted during the analysis of these Balmer lines. We obtained $T_\mathrm{eff} = 6160 \pm 140$\,K, in perfect agreement with the result from photometry. Based on these results, we adopted an effective temperature of $6160 \pm 100$\,K for our analysis. Attempts were also made to derive the effective temperature by comparing the observed colours with calculated colours of the MARCS model atmospheres (Edvardsson et al. 2005, in preparation). We found values of $T_\mathrm{eff}$ from $B-V$, $V-R$ and $V-I$ to range between 6100 K and 6350 K.

\begin{figure}
\begin{center}
\resizebox{\hsize}{!}{\rotatebox{0}{\includegraphics{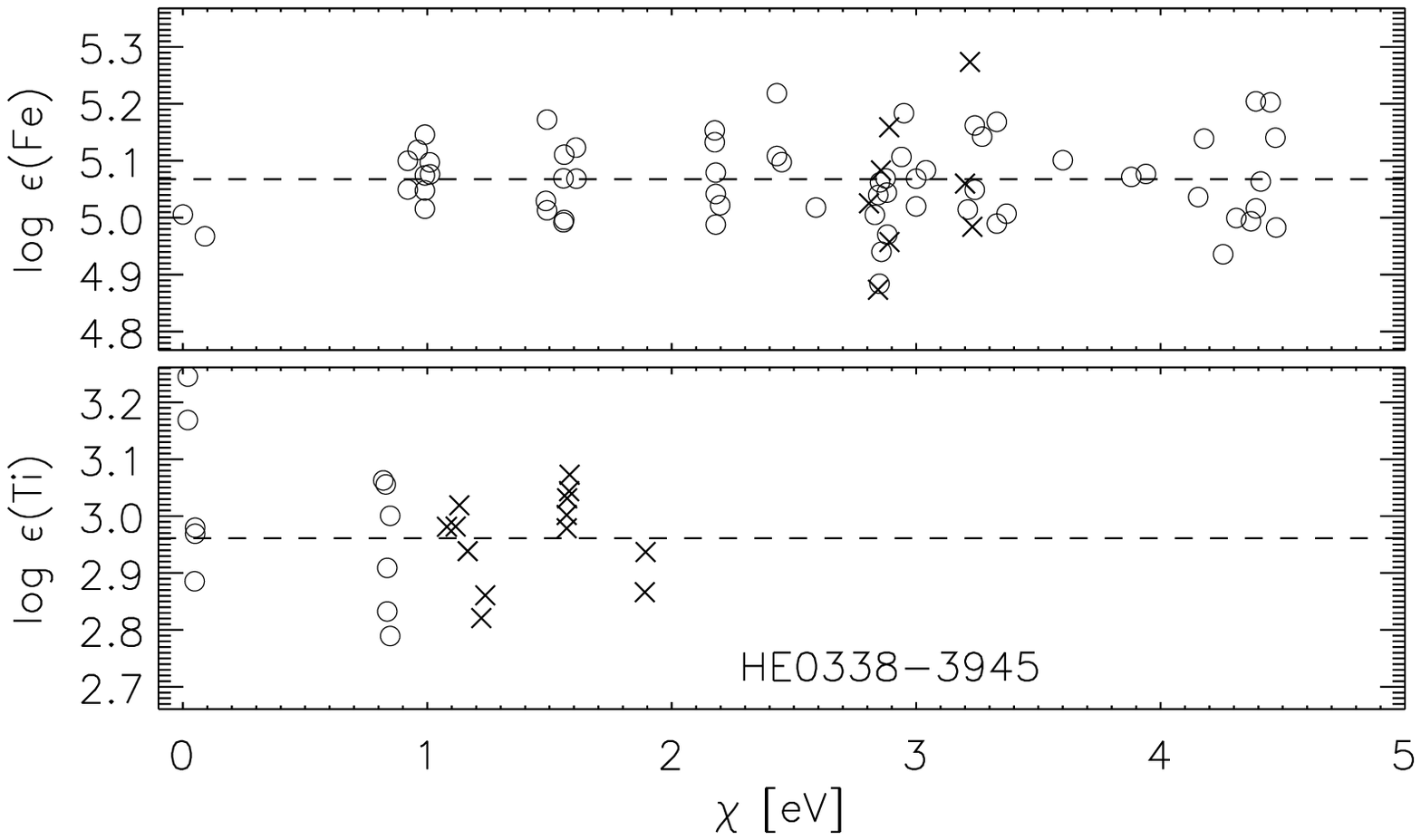}}}
\resizebox{\hsize}{!}{\rotatebox{0}{\includegraphics{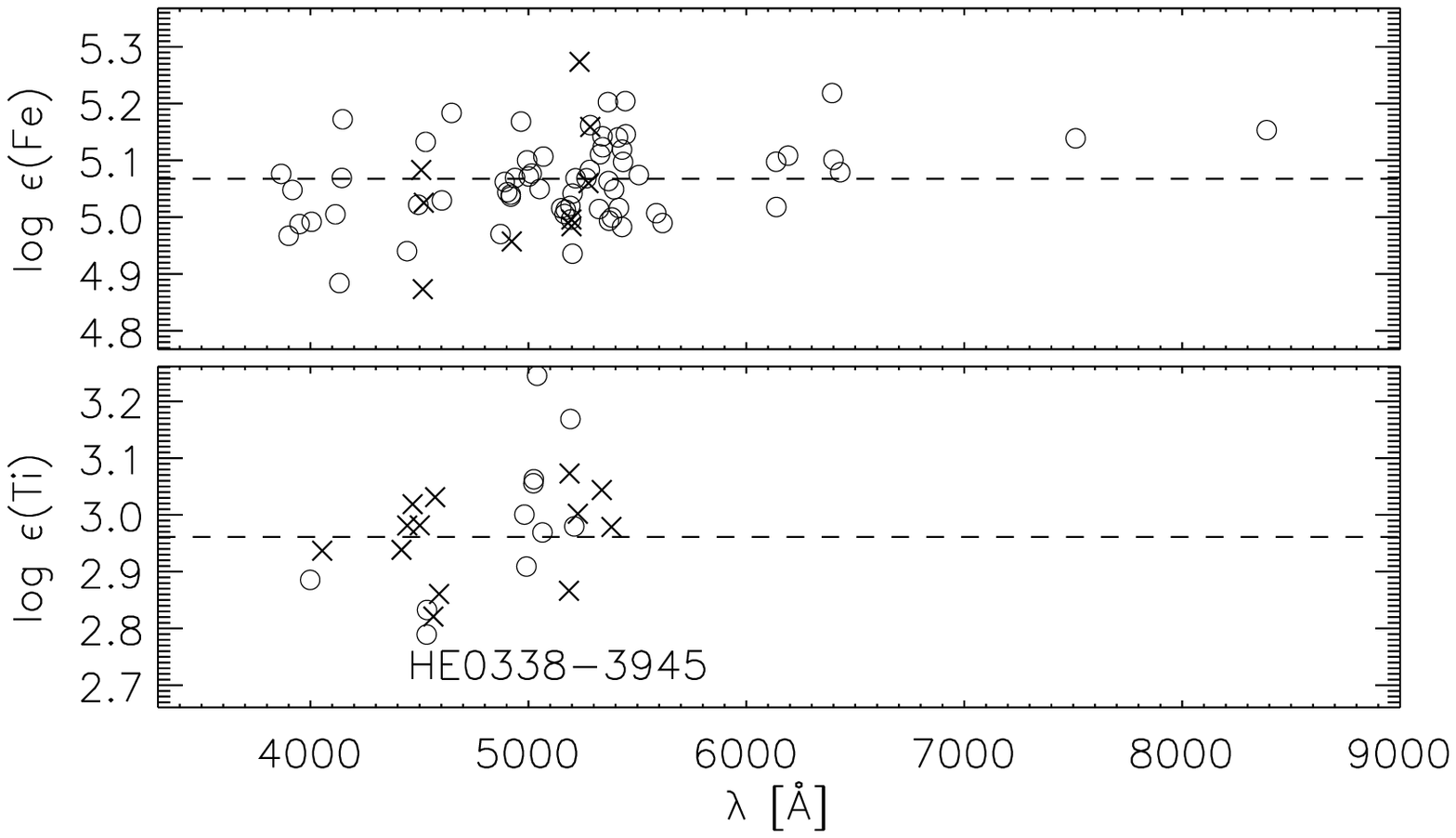}}}
\resizebox{\hsize}{!}{\rotatebox{0}{\includegraphics{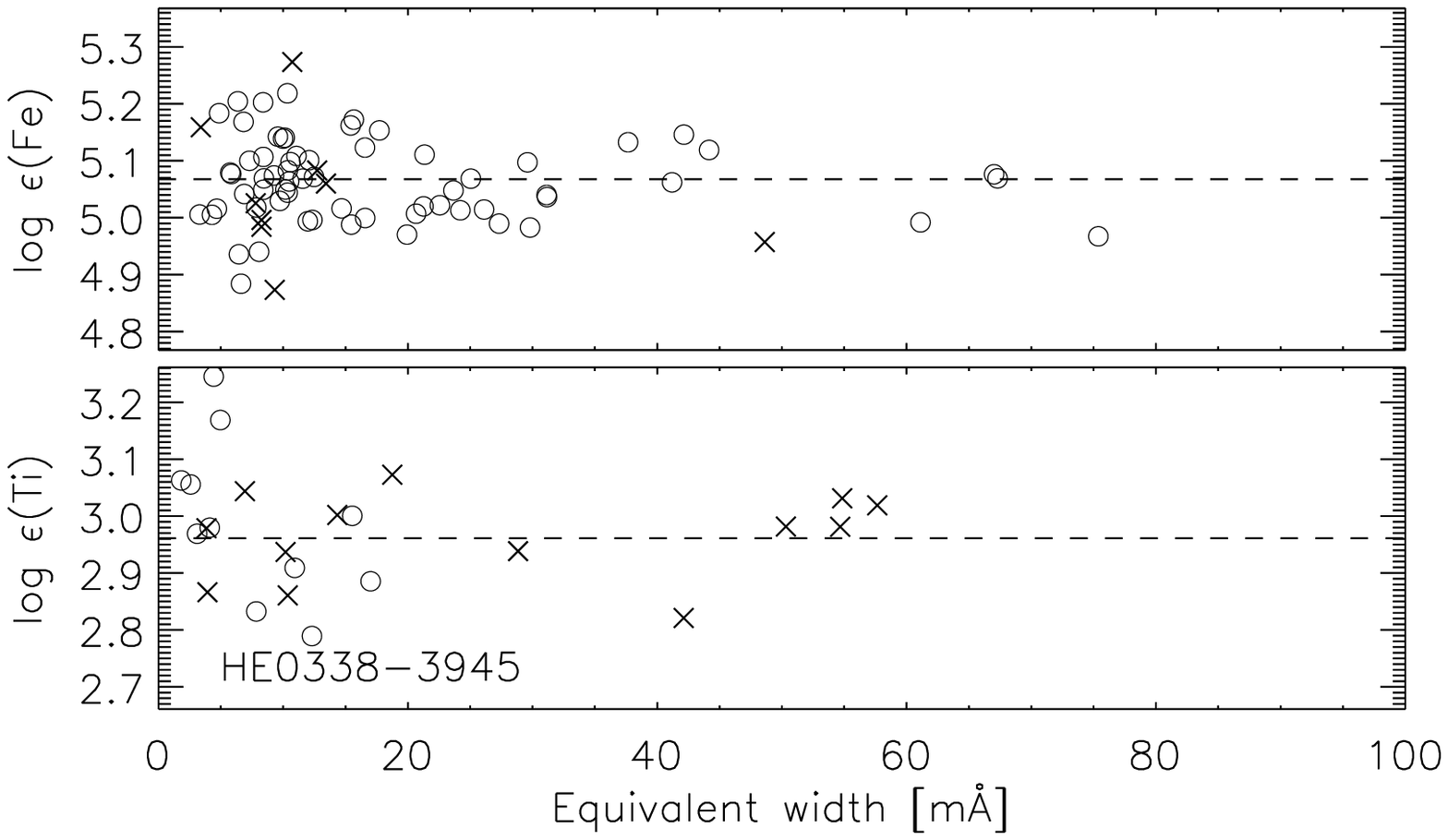}}}
\end{center}
\caption{Plots showing trends of abundances found from individual Fe and Ti lines with excitation potential ({\it top}), wavelength ({\it middle}) and equivalent width ({\it bottom}). Circles and crosses indicate lines from the neutral and singly ionised species respectively, and the dashed lines indicate the determined abundances from the simultaneous best fit to all lines of the element in question.}
\label{fig:equilib}
\end{figure}

A small grid of models in $\log g$ and [Fe/H] for a fixed $T_\mathrm{eff}$ was calculated with the updated Uppsala model atmosphere code MARCS (Gustafsson et al. in preparation). All models used scaled solar abundances with the exception of the alpha-elements, which were enhanced by 0.4\,dex. A microturbulence of 1\,km/s was adopted for the line opacities. This grid was used for an initial analysis of the spectrum, deriving $\log g$ and chemical abundances as described in Sect.\,\ref{program}. The parameters from Paper\,{\sc ii} were used as the initial guess. As C, N and O abundances are significantly enhanced in this star, the grid was then recomputed enhancing these abundances by amounts suggested by the initial analysis. The changes in these abundances led to a warming of the model atmosphere by about 25\,K in the typical line forming regions. A final analysis of the spectrum was then made as described in Sect.\,\ref{program}, giving final model parameters of $T_\mathrm{eff} = 6160 \pm 100$\,K, $\log g = 4.13 \pm 0.33$\,dex, and $\mathrm{[Fe/H]} = -2.42 \pm 0.11$. The final microturbulence $\xi_\mathrm{t}$ was $1.13 \pm 0.22$\,km/s. In Fig.\,\ref{fig:equilib} we show that adopted model is consistent with excitation and ionisation equilibrium for Fe and Ti, and that there are no trends in Fe and Ti abundances with line strength or wavelength. We compared these stellar parameters with a $\mathrm{[Fe/H]} = -2.5$ and 12 Gyr isochrone from Kim et~al.~(\cite{kim2002}), and the result lies near the turnoff but between the positions of the main-sequence and the sub-giant stage. We note that $\log g$ is expected to be underestimated since overionisation of \ion{Fe}{i} will lead to an underestimate of the gravity in our LTE analysis (estimating the pressure in the atmosphere to be lower than it actually is), and thus this star is most likely closer to the main-sequence than the sub-giant branch.

\subsection{Line selection and data}
\label{lines}

The lines used in the abundance analysis were selected through a thorough search and selection process.  All lines in the observed wavelength regions were extracted from the Vienna Atomic Line Data Base (VALD, Kupka et~al.~\cite{kupka1999}).  Predicted equivalent widths were then calculated based on a MARCS model atmosphere with the stellar parameters and abundances of Paper\,{\sc ii}.  For $n$-capture element abundances not derived in Paper\,{\sc ii} we used solar values appropriately scaled to have similar enhancements to other $n$-capture elements. Predicted strengths of C$_\mathrm{2}$, CH and CN features, in the wavelength ranges were also calculated.  For elements with many available lines of reasonable strength, lines with equivalent widths predicted to be smaller than $< 0.1$\,m{\AA} were rejected, and the remaining lines which were predicted not to be overly blended were then scrutinised in the UVES spectra, checking that the line is observed and is indeed unblended. For elements of interest with very few lines, (e.g.\ Ag, Tb) all possible lines were scrutinised.   The lines from the VALD search were complemented with lines of Paper\,{\sc ii}, and Sneden et~al.~(\cite{sneden1996, sneden2003}).  The preliminary line list was then applied to an initial analysis of the spectrum (see Sect.\,3.3), and following visual inspection of the results each line was either adopted in the final line list or rejected.   Lines were rejected if the predicted line from the derived abundance did not agree well with the observed line, with respect to the majority lines of the element.  This generally indicates a blend or erroneous $\log gf$.  A large number of lines were rejected in this way.  Most blends were found to be due to CH and Nd. 

The final line list comprises 621 lines for measuring abundances and 650 blending lines.  Blending lines are modelled in the calculated synthetic spectrum using abundances derived from other lines, but are not themselves used in determining the abundances.  Atomic data has been compiled for the selected lines.  Where several sources of atomic data were available, the data were generally extracted in the following order of priority: Paper\,{\sc ii}, Sneden et~al.~(\cite{sneden2003}), Sneden et~al.~(\cite{sneden1996}), laboratory data from VALD, Jonsell et~al.~(\cite{jonsell2005}), theoretical data from VALD.  In a few cases, we were aware of recent laboratory $f$-values which were preferred, e.g. in the case of Nd.   In general this reflects our preference for well checked data, and for laboratory data over astrophysical or theoretical data. The original sources of the data were traced where possible, and are presented with the final line list in Table 2, which is only available electronically.  The wavelength $\lambda$, excitation potential $\chi$, $\log gf$-value, collisional broadening due to neutral hydrogen, Stark broadening, and radiation damping constant, $\gamma_\mathrm{rad}$, are tabulated in Table 2, along with the spectral windows used in the abundance analysis (see the description of the method in Paper\,{\sc ii}). When applicable and available, data for the isotopic splitting and hyperfine structure (hfs) are given. Blending lines considered are also included in the list. In the Appendix we briefly comment on line selection issues and data sources for each element. 

\subsection{Analysis method} 
\label{program}                               

The spectrum has been analysed using the automated spectrum analysis code based on SME (Valenti \& Piskunov~\cite{valenti1996}) which is described in detail in Paper\,{\sc ii}. Here we summarise the most relevant aspects of the method. The synthetic spectrum was calculated assuming LTE, a 1D plane-parallel geometry for the atmosphere, and Doppler broadening was modelled through the classical microturbulence and macroturbulence parameters. Model parameters were optimised to minimise the $\chi^2$ statistic comparing the synthetic and observed spectra. The effective temperature was derived independently as described in Sect.\,\ref{parametersmodel} and held fixed throughout. The remaining atmospheric parameters ($\log g$, [Fe/H], $\xi_\mathrm{t}$, $v_\mathrm{macro}$) were first determined from an analysis of Fe and Ti lines. These parameters were then adopted, and abundances for all elements determined from appropriate lines considering also blends in the region. As a large number of lines regarded as blends were included in this analysis, a second iteration was performed to ensure that these lines were modelled with appropriate abundances. In some instances where lines were undetectable, we calculated upper limits to the abundances.  We computed 3$\sigma$ upper limits using the same code by finding the abundance necessary to produce a line with an equivalent width three times the 1$\sigma$ measurement error in the equivalent width due to noise that a weak line at the relevant location in the spectrum would have.

One significant upgrade of the code was performed, namely the partition functions from the MOOG code (Sneden~\cite{sneden1973}) have been adopted, which have been significantly updated and corrected in 2002 (see http://verdi.as.utexas.edu/moog.html). Some significant corrections to the previously used polynomial fits to the partition functions of Irwin~(\cite{irwin1981}) have been found, e.g. \ion{Tb}{ii} (Lawler et~al.~\cite{lawler2001b}).
      
Error estimates in all quantities are computed as detailed in Paper\,{\sc ii}. The propagation of errors from stellar parameters, oscillator strengths, observational error and continuum placement are considered in detail. In Paper\,{\sc ii} we also estimated contributions to the error from modelling uncertainties, such as the assumptions of LTE and 1D plane-parallel geometry, as these may contribute to star-to-star scatter in a sample of stars. However, in this case we are examining a single star and this error has not been included, i.e.\ $\sigma_\varepsilon(\mathrm{model}) = 0$. 

As in Paper\,{\sc ii}, we make a distinction between absolute and relative error estimates. The absolute error, $\sigma^{\mathrm{abs}}$, estimates the total uncertainty in a given quantity. The relative error, $\sigma^{\mathrm{rel}}$, is an error where mainly the uncertainty in the line $f$-values ($\sigma_{\log gf}$) has been neglected. Thus, it is appropriate for star-to-star comparison where the same lines and $f$-values are used, as in our homogeneous analysis with HE~2148$-$1247 (see Sect.\,\ref{comparison_cohen}).

Since the procedure globally fits the spectral lines of a given element, an average uncertainty for each element $\sigma_{\log gf}$ is required when computing $\sigma^{\mathrm{abs}}$. As a number of elements studied here were not included in Paper\,{\sc ii}, in Table\,\ref{tab:loggf_unc} we provide an estimated average uncertainty for each element with reference to the original literature, which are adopted for $\sigma_{\log gf}$. In the cases of Ho and Hf, we could find no error estimate for the oscillator strengths and adopted 0.2\,dex as an estimate. 

\setcounter{table}{2}
\begin{table}
\begin{center}
\caption{Assigned average values of $\sigma_{\log gf}$ for each element.}
\label{tab:loggf_unc}
\scriptsize
\begin{tabular}{llllll}
\hline
\hline
Element &  $\sigma_{\log gf}$ & Element &  $\sigma_{\log gf}$ & Element &  $\sigma_{\log gf}$\\
\hline
C & 0.10  & Co & 0.10  & Eu & 0.03 \\
N & 0.10  & Ni & 0.03  & Gd & 0.05 \\
O & 0.02  & Cu & 0.04  & Tb & 0.04 \\
Na & 0.02 & Sr & 0.10  & Dy & 0.04 \\
Mg & 0.07 & Y  & 0.03  & Ho & 0.20 \\
Al & 0.11 & Zr & 0.03  & Er & 0.12 \\ 
Ca & 0.11 & Ag & 0.04  & Tm & 0.02 \\
Sc & 0.04 & Ba & 0.03  & Yb & 0.05 \\ 
Ti & 0.05 & La & 0.03  & Lu & 0.04 \\
V  & 0.05 & Ce & 0.10  & Hf & 0.20 \\
Cr & 0.05 & Pr & 0.04  & Pb & 0.05 \\
Mn & 0.06 & Nd & 0.03  & Th & 0.02 \\
Fe & 0.03 & Sm & 0.05  & U  & 0.05 \\
\hline
\end{tabular}
\end{center}
\end{table} 
                                                       
\section{Results}
\label{resultsA} 

\subsection{Abundance results}
\label{results} 

\begin{figure*}
\centering
\resizebox{\hsize}{!}{\includegraphics*{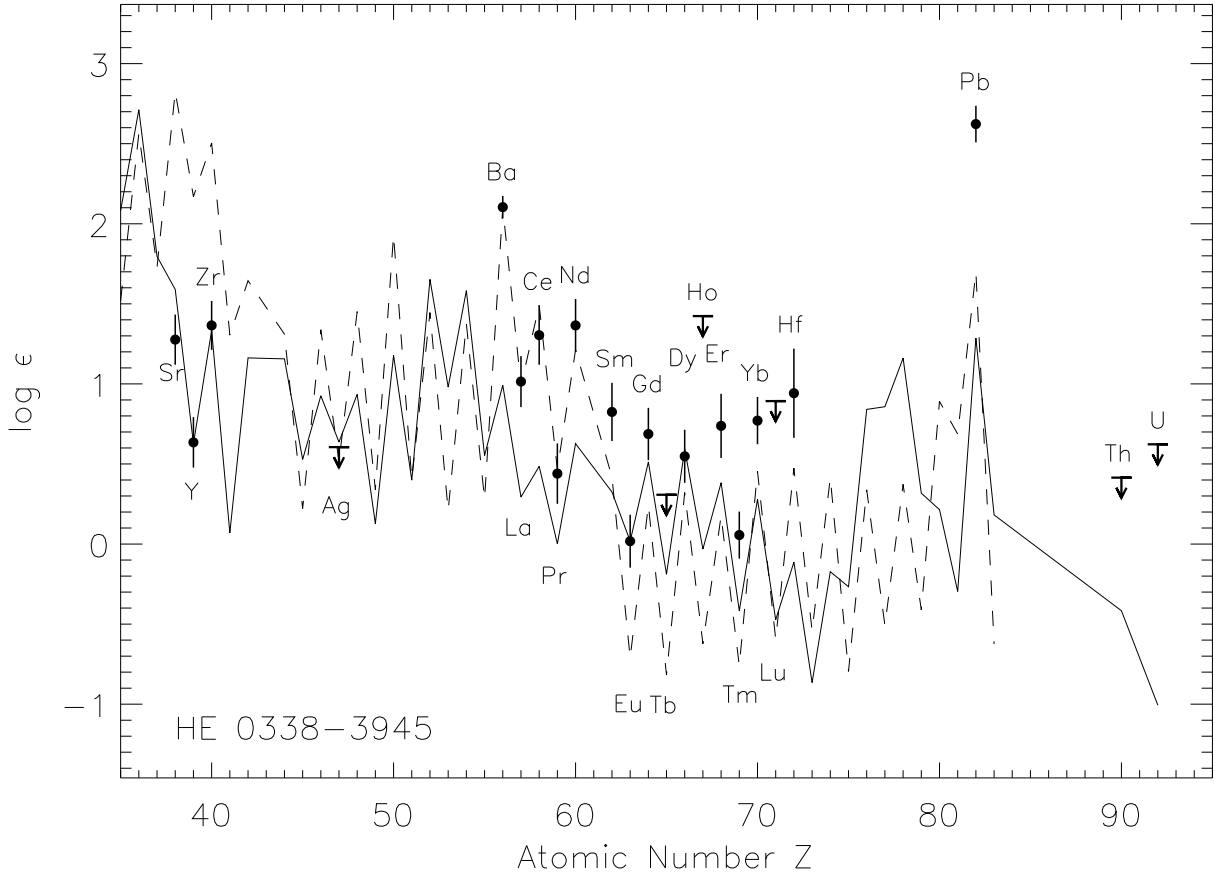}}
\caption[]{
The abundance pattern of HE~0338$-$3945 (full circles and downward arrows) compared with the solar $r$-process pattern (solid line) scaled to the Eu abundance of HE~0338$-$3945, and the solar $s$-process pattern (dotted line) scaled to the Ba abundance of the star. The estimated absolute error bars are shown. The $r$- and $s$-process fractions are from Arlandini et~al.~(\cite{arlandini1999}), except for Th and U which are from Burris et~al.~(\cite{burris2000}).  Note, in the case of upper limits (downward arrows) the plotted horizontal bar marks the upper limit {\em plus} the 1$\sigma$ absolute error estimate in the upper limit.}
\label{FHE0338-3945_rs}
\end{figure*}

We have derived abundances for 33 elements and upper limits for an additional 6 elements. The results of the abundance analysis are given in Table\,\ref{Tabundance} and the results for the $n$-capture elements are displayed in Fig.\,\ref{FHE0338-3945_rs}.  In the table we present the derived 1D LTE abundances together with the estimated relative and absolute errors (see Sect.\,\ref{program}).  In calculating the error estimates, the sensitivity of each abundance to a change in each stellar parameter is calculated, and the results are shown in Table\,\ref{Terrors}.

\begin{table*}
\begin{center}
\caption{
Derived elemental abundances for HE~0338$-$3945. For each element X the LTE abundance is presented as $\log\epsilon_{\mathrm{X}}$ and [X/Fe], together with their relative, $\sigma^{\mathrm{rel}}$,and absolute, $\sigma^{\mathrm{abs}}$, r.m.s.\ error estimates (see text).  N$_{tot}$ gives the number of lines of the element used for abundance determination for each element, and N$_{\mathrm{3}\sigma}$ gives the number of those features classified as 3$\sigma$ detections.}
\label{Tabundance}
\footnotesize
\begin{tabular}{lllrrrrrrrr}
\hline
\hline
\multicolumn{3}{l}{Element} & 
$\log\epsilon_\mathrm{X}$ & 
$\sigma^{\mathrm{rel}}_{\log\epsilon_\mathrm{X}}$ & 
$\sigma^{\mathrm{abs}}_{\log\epsilon_\mathrm{X}}$ & 
[X/Fe] & 
$\sigma^{\mathrm{rel}}_{\mathrm{[X/Fe]}}$ & 
$\sigma^{\mathrm{abs}}_{\mathrm{[X/Fe]}}$ & 
N$_{\mathrm{tot}}$ & 
N$_{\mathrm{3}\sigma}$ \\ 
& & & [dex] & [dex] & [dex] & [dex] & [dex] & [dex] & & \\
\hline 
 6 & C  & (CH)      &     8.08 & 0.09 & 0.15 &     2.13 & 0.06 & 0.15 & -- & --$^1$ \\  
 7 & N  & (CN)      &     6.92 & 0.16 & 0.20 &     1.55 & 0.09 & 0.17 & -- & --$^1$ \\  
 8 & O  & (O I)     &     7.66 & 0.04 & 0.09 &     1.40 & 0.09 & 0.11 &   2 &   1 \\  
11 & Na & (Na I)    &     4.26 & 0.03 & 0.06 &     0.36 & 0.07 & 0.10 &   3 &   2 \\  
12 & Mg & (Mg I)    &     5.45 & 0.02 & 0.09 &     0.30 & 0.09 & 0.14 &   7 &   5 \\  
13 & Al & (Al I)    &     3.16 & 0.09 & 0.15 &  $-$0.88 & 0.03 & 0.13 &   1 &   1 \\  
20 & Ca & (Ca I)    &     4.31 & 0.05 & 0.12 &     0.38 & 0.07 & 0.14 &  10 &   7 \\  
21 & Sc & (Sc II)   &     1.26 & 0.13 & 0.16 &     0.53 & 0.05 & 0.09 &    5 &   2 \\  
22 & Ti & (Ti I+II) &     2.96 & 0.12 & 0.14 &     0.37 & 0.04 & 0.08 &   24 &  11 \\  
23 & V  & (V I+II)  &     1.76 & 0.13 & 0.16 &     0.19 & 0.05 & 0.09 &    5 &   0 \\  
24 & Cr & (Cr I+II) &     3.12 & 0.09 & 0.11 &  $-$0.12 & 0.03 & 0.08 &    8 &   2 \\  
25 & Mn & (Mn I+II) &     2.47 & 0.11 & 0.12 &  $-$0.49 & 0.04 & 0.08 &    5 &   4 \\  
26 & Fe & (Fe I+II) &     5.07 & 0.10 & 0.11 &     0.00 & 0.03 & 0.05 &   69 &  30 \\  
27 & Co & (Co I)    &     2.72 & 0.09 & 0.14 &     0.23 & 0.03 & 0.11 &    4 &   3 \\  
28 & Ni & (Ni I)    &     3.83 & 0.10 & 0.11 &     0.01 & 0.03 & 0.06 &    8 &   5 \\  
29 & Cu & (Cu I)    &     1.02 & 0.12 & 0.12 &  $-$0.75 & 0.06 & 0.08 &    2 &   0 \\  
38 & Sr & (Sr I+II) &     1.28 & 0.08 & 0.16 &     0.74 & 0.03 & 0.14 &    5 &   2 \\  
39 & Y  & (Y II)    &     0.63 & 0.13 & 0.16 &     0.83 & 0.05 & 0.08 &   11 &   6 \\  
40 & Zr & (Zr II)   &     1.37 & 0.13 & 0.15 &     1.20 & 0.04 & 0.08 &   16 &   7 \\  
47 & Ag & (Ag I)    & $<$ 0.45 & 0.15 & 0.16 & $<$ 1.94 & 0.11 & 0.12 &    2 &   0 \\
56 & Ba & (Ba II)   &     2.10 & 0.06 & 0.07 &     2.41 & 0.05 & 0.08 &    3 &   3 \\  
57 & La & (La II)   &     1.01 & 0.13 & 0.16 &     2.28 & 0.04 & 0.08 &    8 &   6 \\  
58 & Ce & (Ce II)   &     1.30 & 0.14 & 0.18 &     2.16 & 0.05 & 0.12 &   15 &   8 \\  
59 & Pr & (Pr II)   &     0.44 & 0.16 & 0.19 &     2.16 & 0.07 & 0.11 &    4 &   0 \\  
60 & Nd & (Nd II)   &     1.37 & 0.14 & 0.16 &     2.30 & 0.05 & 0.08 &   34 &  15 \\  
62 & Sm & (Sm II)   &     0.82 & 0.15 & 0.18 &     2.25 & 0.06 & 0.10 &    4 &   1 \\  
63 & Eu & (Eu II)   &     0.02 & 0.14 & 0.17 &     1.94 & 0.05 & 0.09 &    3 &   3 \\  
64 & Gd & (Gd II)   &     0.69 & 0.13 & 0.16 &     2.00 & 0.05 & 0.09 &    8 &   0 \\  
65 & Tb & (Tb II)   & $<$ 0.13 & 0.16 & 0.18 & $<$ 2.66 & 0.09 & 0.11 &    1 &   0 \\
66 & Dy & (Dy II)   &     0.55 & 0.14 & 0.17 &     1.84 & 0.05 & 0.09 &   16 &   3 \\  
67 & Ho & (Ho II)   & $<$ 1.16 & 0.15 & 0.27 & $<$ 3.33 & 0.08 & 0.23 &    1 &   0 \\
68 & Er & (Er II)   &     0.74 & 0.14 & 0.20 &     2.24 & 0.05 & 0.15 &    3 &   1 \\  
69 & Tm & (Tm II)   &     0.06 & 0.13 & 0.15 &     2.49 & 0.05 & 0.08 &    4 &   0 \\  
70 & Yb & (Yb II)   &     0.77 & 0.13 & 0.15 &     2.12 & 0.07 & 0.09 &    2 &   2 \\  
71 & Lu & (Lu II)   & $<$ 0.72 & 0.15 & 0.17 & $<$ 3.09 & 0.09 & 0.12 &    1 &   0 \\
72 & Hf & (Hf II)   &     0.94 & 0.17 & 0.28 &     2.49 & 0.10 & 0.23 &    1 &   1 \\  
82 & Pb & (Pb I)    &     2.62 & 0.10 & 0.11 &     3.10 & 0.04 & 0.08 &    2 &   1 \\  
90 & Th & (Th II)   & $<$ 0.23 & 0.17 & 0.19 & $<$ 2.57 & 0.08 & 0.11 &    1 &   0 \\
92 & U  & (U II)    &$< -$0.11 & 0.38 & 0.74 & $<$ 2.82 & 0.46 & 0.80 &    1 &   0 \\
\hline
\multicolumn{11}{l}{($^1$) The detection exceeds 3$\sigma$ for the molecular bands seen as single features.} \\
\end{tabular}
\end{center}
\end{table*}

\begin{table}
\begin{center}
\caption{
The change in the logarithmic abundances due to a change of the fundamental parameters by amounts corresponding approximately to their uncertainties.}
\label{Terrors}
\scriptsize
\begin{tabular}{lrrr}
\hline 
\hline 
 Element & $T_\mathrm{eff}+100$\,K & $\log g+0.3$\,dex  & $\xi_\mathrm{t}+0.2$\,km/s  \\
         & [dex]     & [dex] & [dex]        \\
\hline 
   C   	  &    0.15  &  $-0.09$ &  $ -0.01$  \\
   N   	  &    0.23  &  $-0.09$ &  $  0.00$  \\
   O   	  &  $-0.06$ &    0.10  &  $  0.00$  \\
  Na   	  &    0.06  &  $-0.04$ &  $ -0.03$  \\
  Mg   	  &    0.06  &  $-0.06$ &  $ -0.01$  \\
  Al   	  &    0.11  &  $-0.03$ &  $ -0.03$  \\
  Ca   	  &    0.06  &  $-0.02$ &  $  0.00$  \\
  Sc   	  &    0.05  &    0.10  &  $ -0.03$  \\
  Ti   	  &    0.05  &    0.08  &  $ -0.03$  \\
   V   	  &    0.05  &    0.09  &  $  0.00$  \\
  Cr   	  &    0.09  &  $-0.01$ &  $ -0.04$  \\
  Mn   	  &    0.10  &    0.01  &  $ -0.01$  \\
  Fe   	  &    0.07  &    0.03  &  $ -0.02$  \\
  Co   	  &    0.09  &    0.01  &  $ -0.01$  \\
  Ni   	  &    0.09  &    0.01  &  $ -0.04$  \\
  Cu   	  &    0.10  &    0.00  &  $ -0.01$  \\
  Sr   	  &    0.08  &  $-0.02$ &  $ -0.09$  \\
   Y   	  &    0.05  &    0.09  &  $ -0.04$  \\
  Zr   	  &    0.06  &    0.09  &  $ -0.02$  \\
  Ag$^1$  &    0.11  &    0.01  &  $ -0.01$  \\
  Ba      &    0.08  &  $-0.03$ &  $ -0.02$  \\
  La      &    0.05  &    0.10  &  $ -0.01$  \\
  Ce      &    0.06  &    0.10  &  $ -0.01$  \\
  Pr      &    0.06  &    0.12  &  $  0.00$  \\
  Nd      &    0.06  &    0.10  &  $ -0.02$  \\
  Sm      &    0.07  &    0.11  &  $  0.00$  \\
  Eu      &    0.06  &    0.11  &  $  0.00$  \\
  Gd      &    0.05  &    0.10  &  $  0.00$  \\
  Tb$^1$  &    0.07  &    0.09  &  $ -0.01$  \\
  Dy      &    0.07  &    0.09  &  $ -0.02$  \\
  Ho$^1$  &    0.05  &    0.11  &  $  0.00$  \\
  Er      &    0.06  &    0.10  &  $ -0.01$  \\
  Tm      &    0.05  &    0.09  &  $ -0.01$  \\
  Yb      &    0.07  &    0.06  &  $ -0.03$  \\
  Lu$^1$  &    0.05  &    0.09  &  $  0.00$  \\
  Hf      &    0.08  &    0.09  &  $ -0.06$  \\
  Pb      &    0.11  &  $-0.02$ &  $ -0.01$  \\
  Th$^1$  &    0.09  &    0.10  &  $ -0.01$  \\
   U$^1$  &    0.35  &  $-0.80$ &  $ -0.01$  \\
\hline
\multicolumn{4}{l}{($^1$) Effects on the upper limits are shown.} \\
\end{tabular}
\end{center}
\end{table}

As has been described above, this analysis is based on the assumptions of 1D plane-parallel geometry and LTE.  Such a 1D LTE analysis is still the most usual way to estimate the chemical composition of stars.  However, in recent decades, more and more studies of the effects of deviations from these traditional assumptions have been carried out, and these studies have shown that 1D LTE analysis may yield abundances far from the ``true'' values. This is particularly true for metal-poor stars, which according to 3D calculations are on average much cooler in the line forming layers than predicted by 1D model atmospheres (Asplund~\cite{asplund2004}). The abundance effects may be as large as $-$0.5\,dex for minority species and low excitation lines, and even more for molecular lines. Also severe NLTE effects may arise due to the steeper temperature structures and temperature inhomogeneities (Asplund~\cite{asplund2005}).  

While this analysis is a 1D LTE analysis, and the 1D LTE abundances will be discussed in the paper, it is of interest to examine the estimated effects of 3D and NLTE, and these should be born in mind in interpretation of the abundances.  For the purposes of our following discussion, we define the estimated effects:
\begin{eqnarray}
\Delta                 & = & \log\epsilon_{\mathrm{NLTE,3D}} - \log\epsilon_{\mathrm{LTE,1D}}, \\
\Delta_{\mathrm{NLTE}} & = & \log\epsilon_{\mathrm{NLTE,1D}} - \log\epsilon_{\mathrm{LTE,1D}}, \\
\Delta_{\mathrm{3D}}   & = & \log\epsilon_{\mathrm{LTE,3D}} - \log\epsilon_{\mathrm{LTE,1D}}.
\end{eqnarray}
In the absence of complete line formation calculations in 3D-NLTE, of which there are presently very few, the effects for the 1D-NLTE and 3D-LTE cases are discussed.  Note that $\Delta$ is generally different from the sum of $\Delta_{\mathrm{NLTE}}$ and $\Delta_{\mathrm{3D}}$.  In Table\,\ref{Tnlte3d} we present estimates of the 3D and NLTE effects for this star and the spectral features employed based on results from the literature.  The values are in a number of cases uncertain.  In particular the estimates of NLTE effects are often plagued by uncertainties in collision rates. Note that a range of 3D effects are often quoted as the effects on individual lines may vary. 

\begin{table}
\tabcolsep1.5mm
\begin{center}
\caption{
Estimated effects of NLTE and 3D on abundances from literature and adapted to the parameters of HE~0338$-$3945. The numbers are in many cases highly uncertain; see text for further discussion. References: 
(a) Asplund~(\cite{asplund2004}),
(b) Asplund~(\cite{asplund2005}),
(c) Asplund \& Garc{\'i}a P{\'e}rez~(\cite{asplund2001}),
(d) Baum{\"u}ller et al.~(\cite{baumuller1998}),
(e) Baum{\"u}ller \& Gehren (\cite{baumuller1997}),
(f) Collet et~al.~(\cite{collet2005}),
(g) Gehren et~al.~(\cite{gehren2004}),
(h) Korn \& Mashonkina (2006, in preparation)
(i) Nissen et~al.~(\cite{nissen2002}).}
\label{Tnlte3d}
\scriptsize
\begin{tabular}{llcrcccc}
\hline
\hline
\multicolumn{2}{l}{Element} & &
$\Delta_{\mathrm{NLTE}}$ & 
Ref.$_{\mathrm{NLTE}}$ & &
$\Delta_{\mathrm{3D}}$ &
Ref.$_{\mathrm{3D}}$ \\ 
& & & [dex] & & & [dex] & \\     
\hline 
 6 & C (CH)            &  &         &      &  &  $-$0.5 to $-$0.3 & a,c \\  
 7 & N (CN)            &  &         &      &  &		   $-$0.5 & see text \\  
 8 & O                 &  & $-$0.1  & i    &  &  $-$0.1	          & b \\  
11 & Na                &  & $-$0.4  & d    &  &  $-$0.1 to +0.1   & a \\  
12 & Mg                &  &   +0.1  & g    &  &  $-$0.5 to $-$0.3 & a \\  
13 & Al                &  &   +0.5  & e    &  &  $-$0.5 to $-$0.3 & a \\  
20 & Ca                &  &   +0.2  & h    &  &  $-$0.3 to $-$0.1 & a \\  
26 & Fe (\ion{Fe}{i})  &  &   +0.6  & f    &  &  $-$0.5 to $-$0.3 & a \\  
26 & Fe (\ion{Fe}{ii}) &  &   +0.2  & b    &  &  $-$0.1 to +0.1   & a \\  
38 & Sr                &  &   +0.3  & b    &  &  $-$0.5 to $-$0.3 & a \\  
56 & Ba                &  &   +0.2  & b    &  &  $-$0.5 to $-$0.1 & a \\  
63 & Eu                &  &   +0.1  & b    &  &		          &   \\  
\hline
\end{tabular}
\end{center}
\end{table}

Below, we discuss the results and relevant issues in the analysis for each element in turn.  Elements not discussed individually have a significant number of unblended lines with adequate $S/N$ for a clear detection were available, the analysed lines were all well fit by a single abundance using our adopted model, no NLTE or 3D corrections were available, and the abundance was unremarkable with respect to ``normal'' metal-poor stars (see Sect.~\ref{comparison_pop2}).  Note that line selection and atomic data for each element are discussed in the Appendix.

\paragraph{Carbon:} HE~0338$-$3945 is carbon-rich and the two strong bands and two lines of CH used in the abundance analysis were well fitted.  The abundance is, however, uncertain, due to the strong temperature sensitivity of the CH molecule formation (see Table~\ref{Terrors}). In addition, preliminary calculations suggest that $\Delta_{\mathrm{3D}}$ may amount to approximately $-$0.5 to $-$0.3\,dex (Asplund \& Garc{\'i}a~P{\'e}rez~\cite{asplund2001}, Asplund~\cite{asplund2004}).

\begin{figure}
\centering
\resizebox{\hsize}{!}{\includegraphics*{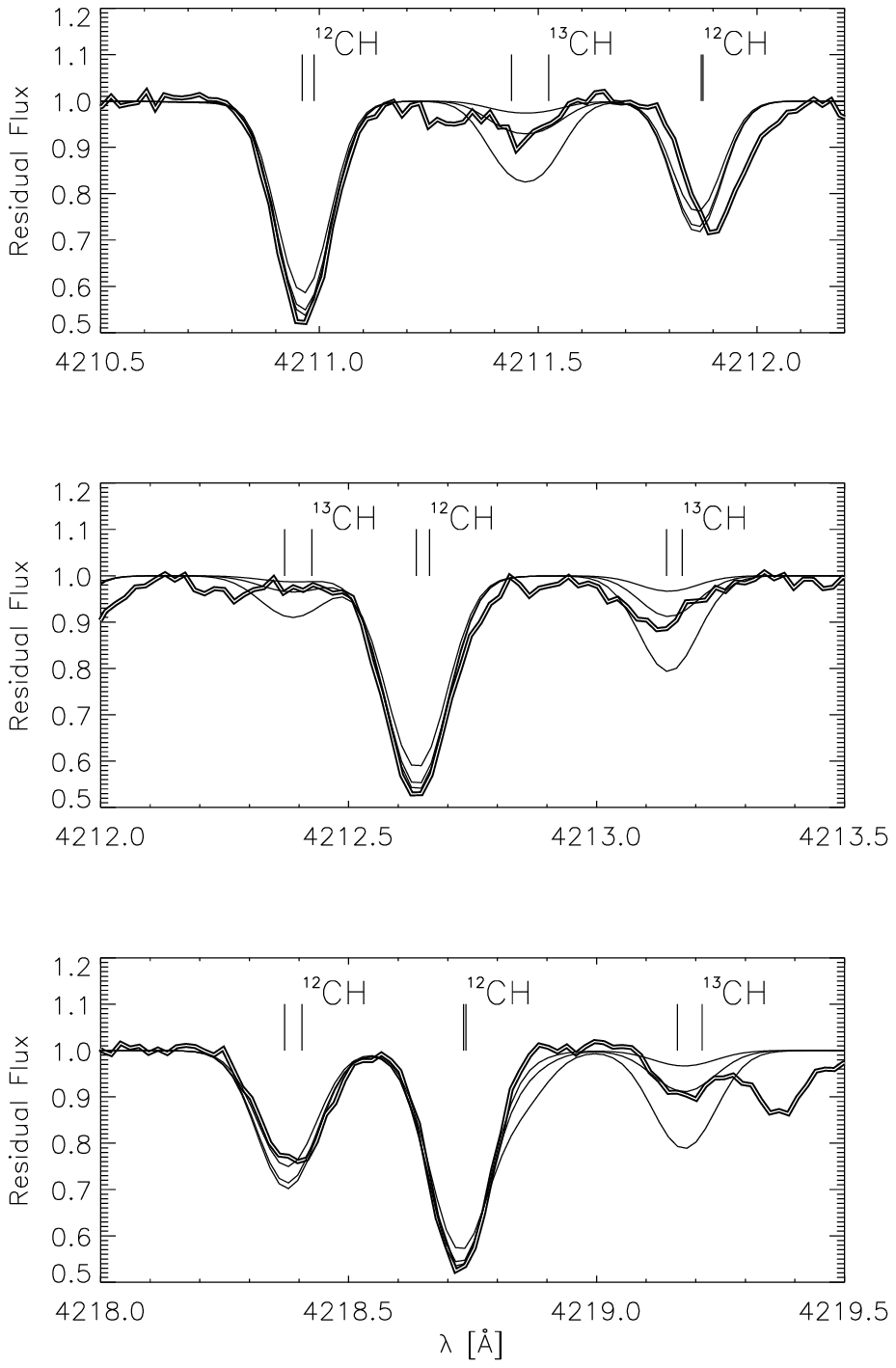}}
\caption[]{
Comparison of observed and synthetic spectra for $^{13}$CH features. The double line shows the observed spectrum. The thin lines show synthetic spectra for isotope ratios $^{12}$C/$^{13}$C=3, 10, and 30.}
\label{F13CH}
\end{figure}

The isotopic ratio $^{12}$C/$^{13}$C was determined from reasonably well isolated $^{13}$CH features between 4210 and 4225\,\AA. The comparisons of observed and synthetic spectra are shown in Fig.\,\ref{F13CH}, where the total derived C abundance derived is adopted. We find $^{12}$C/$^{13}$C$\sim 10$.

\paragraph{Nitrogen:} The N abundance was derived based on bands of the CN molecule.  The abundance is uncertain, first because it is both highly temperature sensitive (see Table~\ref{Terrors}) and dependent on the C abundance, and secondly because it is probably subject to 3D effects. The value of $\Delta_{\mathrm{3D}}$ for CN is expected to be lower than for NH, which may be as much as $-$0.6 to $-$0.9\,dex for turn-off stars (Asplund~\cite{asplund2005}). A reasonable estimate of the effect may amount to $-$0.5\,dex.  Note, the NH A$\Pi-$XZ band at 3360\,{\AA} gave an abundance consistent with the CN result, but was disregarded in the final analysis due to poor $S/N$ and a large degree of blending. 

\paragraph{Oxygen:} Only two of the three lines of the infrared \ion{O}{i} triplet at 7773\,{\AA} were used in the analysis, as the 7775\,{\AA} line was affected by a reduction artefact.  The [\ion{O}{i}] line at 6300\,{\AA} could not be detected.  A value of $\Delta_{\mathrm{NLTE}}\sim-0.1$\,dex has been adopted for stars with similar fundamental parameters by Nissen et~al.~(\cite{nissen2002}). $\Delta_{\mathrm{3D}}$ is estimated to be approximately $-$0.1\,dex by Asplund~(\cite{asplund2005}). 

\paragraph{Sodium:} The derived Na abundance was dominated by the \ion{Na}{i} D lines at 5890\,{\AA}, and is determined quite precisely as seen from the small estimated errors in Table\,\ref{Tabundance}.  However, the true Na abundance may be lower due to NLTE effects. From interpolation in Table~2 of Baum{\"u}ller et~al.~(\cite{baumuller1998}) we estimate $\Delta_{\mathrm{NLTE}}\sim-0.4$\,dex.  The 3D effect seems not to be severe for sodium (Asplund~\cite{asplund2004}).

\paragraph{Magnesium:} The lines of \ion{Mg}{i} were well fit, and the error in the abundance estimated to be quite small. However, it is affected by both NLTE and 3D effects.  According to Asplund~(\cite{asplund2005}) the NLTE effect on the abundance from the \ion{Mg}{i} line at 5172\,{\AA} is around $-$0.2\,dex, while for other \ion{Mg}{i} lines the effect is approximately +0.2\,dex.  We have used a total of 7 lines, and so $\Delta_{\mathrm{NLTE}}$ should be positive.  This is supported by Gehren et~al.~(\cite{gehren2004}) who show results for 7 lines, although not exactly the same set of lines as used here. The resulting value of $\Delta_{\mathrm{NLTE}}$ for stars with parameters close to HE~0338$-$3945 is $\sim +0.1$\,dex.  Asplund~(\cite{asplund2004}) estimate that $\Delta_{\mathrm{3D}}$ may be $-$0.5 to $-$0.3\,dex for the low excitation lines.  

\paragraph{Aluminium:} The abundance of Al was derived using only the strong resonance line of \ion{Al}{i} at 3961\,{\AA}, and gives a clearly sub-solar value relative to Fe as is the norm for metal-poor stars (see Sect.\,\ref{comparison_pop2}). $\Delta_{\mathrm{NLTE}}$ could, according to an interpolation made in Table~1 of Baum{\"u}ller \& Gehren (\cite{baumuller1997}), be as high as +0.5\,dex.  According to Asplund~(\cite{asplund2004, asplund2005}) $\Delta_{\mathrm{3D}}$ may amount to a similar value but of opposite sign for low excitation lines.

\paragraph{Calcium:} The calcium abundance was determined from lines of \ion{Ca}{i}. A value of  $\Delta_{\mathrm{NLTE}} \sim +0.2$\,dex was adopted from recent work of Korn \& Mashonkina (2006, in preparation), and according to Asplund~(\cite{asplund2004}) the $\Delta_{\mathrm{3D}}$ may amount to $-$0.3 to $-$0.1\,dex.

\paragraph{Scandium:} The abundance of scandium was determined using 5 \ion{Sc}{ii} lines. The abundance seems to be high in comparison to normal metal-poor stars (see Sect.\,\ref{comparison_pop2}).  The two bluest lines are strongest and dominate the fitting, and are thus well fit.  The fit does, however, seem to overestimate the strength of the three weak lines redward of 5000~\AA.  This may suggest that the blue lines are affected by unknown or poorly modelled blends.  We compared the [Sc/Fe] and [C/Fe] abundances from Paper\,{\sc ii}, and found that while the results are uncorrelated for $\mathrm{[C/Fe]} \la 1.5$, for $\mathrm{[C/Fe]} \ga 1.5$ there is a trend to high [Sc/Fe] with high [C/Fe].  Thus, our Sc abundance may be overestimated. 

\paragraph{Vanadium:} Five weak lines in the blue were used to derive the abundance, the best being only detected at the $1.8\sigma$ level. However, all lines of both species, \ion{V}{i} and \ion{V}{ii}, are well modelled and we consider this a reliable detection.

\paragraph{Manganese:} Three lines of \ion{Mn}{ii} and two of \ion{Mn}{i} were analysed. The $S/N$ was significantly better at the \ion{Mn}{i} lines than at the \ion{Mn}{ii} lines, and thus the \ion{Mn}{i} lines dominated the fit.  However, the derived abundance did not reproduce the \ion{Mn}{ii} lines well, the model prediction being significantly weaker than the observed by a consistent amount for all three lines. The discrepancy may be due to overionisation of \ion{Mn}{i}.  Such a discrepancy has also been noticed by Johnson~(\cite{johnson2002}).

\paragraph{Iron:} 
A total of 61 \ion{Fe}{i} and 8 \ion{Fe}{ii} lines were analysed. In Table\,\ref{Tnlte3d} we give estimates for $\Delta_{\mathrm{3D}}$ and $\Delta_{\mathrm{NLTE}}$ for each species separately.  As our Fe abundance is dominated by the numerous \ion{Fe}{i} lines the effects on \ion{Fe}{i} should dominate. Note, however, that the effects of NLTE are presently highly uncertain due to uncertainty in inelastic collision rates due to hydrogen atoms.  For example, Korn~et~al.~(\cite{korn2003}) found much smaller NLTE effects when astrophysically calibrating these collision rates.  As noted in Sect.~\ref{parametersmodel} NLTE effects on \ion{Fe}{i} are expected to lead to underestimation of $\log g$ in LTE, and this would affect all other elements.

\paragraph{Copper:} The Cu abundance was derived from two lines in the ultraviolet. Both lines were clearly seen and identified, and well modelled with a single abundance.  However, the low $S/N$ ratio in this spectral region meant that the strongest line was detected only at the $2.9\sigma$ level, and continuum placement was uncertain. Despite this, we regard it as a definite detection.

\paragraph{Strontium, yttrium, zirconium:} Sr, Y and Zr, are strongly overabundant compared to normal metal-poor stars (see Sect.\,\ref{comparison_pop2}).  Interestingly, the LTE abundances of these elements lie below the solar $s$-process abundance pattern as normalised to Ba (see Fig.\,\ref{FHE0338-3945_rs}), and curiously enough, at least Y and Zr fit the $r$-process abundance pattern normalised to Eu.  Strontium is estimated to have $\Delta_{\mathrm{NLTE}}\sim +0.3$\,dex for the lines at 4077, 4161 and 4215\,{\AA} in metal-poor turn-off stars (Asplund \cite{asplund2005}). We analysed two more weak lines, one with low $S/N$.  The stronger lines at higher $S/N$ have high weight in the abundance calculation, and therefore this estimate for  $\Delta_{\mathrm{NLTE}}$ is appropriate.  $\Delta_{\mathrm{3D}}$ is approximately $-0.3$ to $-0.5$ (Asplund~\cite{asplund2004}). 

\paragraph{Silver:} The upper limit to the silver abundance is estimated from two \ion{Ag}{i} lines, both blended and of roughly equal intrinsic strength.  The upper limit is sufficiently low to be inconsistent with the scaled solar $r$-process abundance pattern (see Fig.\,\ref{FHE0338-3945_rs}). This may reflect overionisation of \ion{Ag}{i} (see Sect.\,\ref{discussion}).

\paragraph{Barium:} Ba is strongly overabundant compared to normal metal-poor stars (see Sect.\,\ref{comparison_pop2}). This abundance was derived using three \ion{Ba}{ii} lines: two weak and the strong resonance line. The resonance line at 4554\,{\AA} has an estimated value of $\Delta_{\mathrm{NLTE}}$ of at least +0.2\,dex (Asplund~\cite{asplund2005}).  $\Delta_{\mathrm{3D}}$ is of the opposite sign, and is estimated at $-$0.5 to $-$0.1\,dex (Asplund~\cite{asplund2004}). 

\paragraph{Lanthanum:} Lanthanum is also strongly overabundant in this star.  Eight lines of \ion{La}{ii} of varying strength were employed, the majority of which were well fit by a single abundance. However, two lines, 4920 and 4921\,{\AA} were not well fit, the observed lines being stronger than the predicted lines for the derived abundance. We note that for these lines we had incomplete hfs data (see Appendix), and this could possibly lead to the line strengths being underestimated.

\paragraph{Praseodymium, Neodymium, Samarium:} Four weak lines were employed to derive the abundance of Pr, the best being only detected at the $1.6\sigma$ level. However, all 4 lines were well modelled, and thus we consider this a reliable detection. A total of 34 mostly weak lines were used to derive the Nd abundance, only slightly less than half of the lines being detected at the 3$\sigma$ level. For Sm, four weak lines of \ion{Sm}{ii} were used. 

\paragraph{Europium:} The abundance of Eu is enhanced compared to typical metal-poor stars (see Sect.\,\ref{comparison_pop2}). The abundance may be even higher, as $\Delta_{\mathrm{NLTE}}$ is probably greater than +0.1 (Asplund~\cite{asplund2005}). No estimate of $\Delta_{\mathrm{3D}}$ is available, but see the discussion regarding abundance ratios at the end of this section.

\paragraph{Gadolinium:} Eight weak lines are used, the best being only a $1.8\sigma$ detection.  However, all eight lines are consistently modelled, and we class this as a reliable detection. 

\paragraph{Terbium:} No line was detected, and thus an upper limit was derived from single line of \ion{Tb}{ii}.

\paragraph{Holmium:} No lines are detected, and thus a $3\sigma$ upper limit was derived. Due to strong blending at the strongest lines of Ho, we were forced to use a rather weak line and thus the derived upper limit is rather high and a weak constraint.

\paragraph{Erbium, Thulium:} The abundance of Er was based on three weak \ion{Er}{ii} lines. For Tm, 4 weak lines were employed, the best being only detected at the $1.6\sigma$ level. However, the lines of these elements were well modelled, and thus we consider these as reliable detections. Both elements have abundances significantly above the scaled solar $s$- and $r$-process patterns (see Fig.\,\ref{FHE0338-3945_rs}).

\paragraph{Lutetium:} No lines were detected, and thus a $3\sigma$ upper limit was derived. The employed line region is somewhat blended, but we regard the derived upper limit as reliable. 

\paragraph{Hafnium:} Only the resonance line in the ultraviolet was available, but a definite detection was made. 

\paragraph{Lead:} The Pb abundance is significantly above the scaled $s$-process abundance pattern in Fig.\,\ref{FHE0338-3945_rs}. Pb has an ionization energy of 7.4\,eV, and the two lines observed are both \ion{Pb}{i} lines. Here, overionisation is a possibility, which would lead to an underestimated Pb abundance and e.g.\ Pb/Ba ratio,  meaning that this ratio, already remarkably high in the LTE analysis, may be even higher. 

\paragraph{Thorium:} The $3\sigma$ upper limit was based on a single \ion{Th}{ii} line at 4019.129\,{\AA}. The region contains a number of blends, but a reliable upper limit was possible to derive.

\paragraph{Uranium:} No lines are detected, so a $3\sigma$ upper limit has been derived. The region is significantly blended, and the reliability of the derived upper limit is dependent on the correct modelling of these blends.

\paragraph{Some specific abundance ratios:}
We shall finally comment on the accuracy of some important abundance ratios, to be further discussed below. The similarity of Sr, Y, Zr and Ba in terms of ionisation energies and atomic structure seems to suggest that the effects of NLTE should be limited on Ba/Sr, Ba/Y and Ba/Zr. Also, the most important lines in the analysis have low excitation energies -- the Ba abundance determination is dominated by the wings of the 4554\,{\AA} resonance line, while the other two lines with 2.5 and 2.7\,eV excitation energies are less significant.  Thus, the temperature sensitivity of the important line strengths are roughly similar, which is reflected in similar temperature sensitivities of the abundances in Table\,\ref{Terrors}. This should lead to similar sensitivities to 3D effects.  Similarly, the effects of NLTE and 3D on abundance ratios of these elements relative to Eu may be expected to be small.  

\subsection{Binarity}

Radial velocities were measured from a series of \ion{Fe}{i} lines with accurate laboratory wavelengths between 4000 and 5000\,\AA, using the snapshot spectrum observed on 15 October 2002, and the observations from December 2002 which comprise the high quality spectrum analysed in this work. No difference in radial velocity shift was observed between the different exposures taken in December 2002. The barycentric radial velocities measured from the snapshot spectrum and the co-added high quality spectrum were $177.7 \pm 0.8$\,km/s and $177.9 \pm 0.5$\,km/s respectively, where the quoted errors are the standard deviations of the results from different lines; errors due to the wavelength calibration uncertainties are not included.  We therefore presently have no evidence for the binarity of HE~0338-3945.  A recent study by Lucatello et al.~(\cite{lucatello2005}) suggests all CEMP stars are multiple systems. 

\section{Comparisons}
\label{comparisons}

\subsection{Comparison with the results of Paper\,{\sc ii}}
\label{comparison_paper2}

\begin{figure}
\begin{center}
\resizebox{\hsize}{!}{\rotatebox{0}{\includegraphics{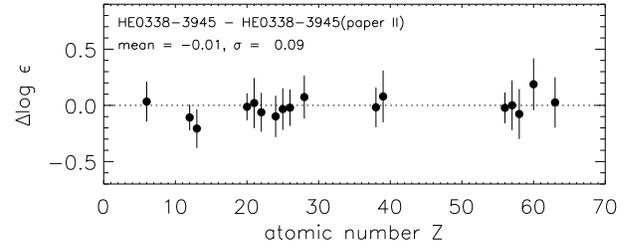}}}
\end{center}
\caption{Difference between the abundances for HE~0338$-$3945 from this work and those from Paper\,{\sc ii}. }
\label{fig:compare_heres2}
\end{figure}

As has been mentioned, a snapshot spectrum of this star was analysed in Paper\,{\sc ii}. The snapshot spectrum of HE~0338$-$3945, compared to the spectrum employed in this work, is of relatively low quality with $S/N \sim 47$ and $R \sim 20000$ and has significantly less wavelength coverage, $\lambda = 3760$--4980\,{\AA}. The snapshot spectrum was reduced by the UVES pipeline, while the spectrum analysed here was reduced as described in Sect.\,\ref{observations}. However, the spectra have been analysed in a similar manner, and it is of interest to check if the results are consistent. We have already seen in Sect.\,\ref{analysis} that the results for the stellar parameters are in good agreement. Figure~\ref{fig:compare_heres2} compares the abundances where available from Paper\,{\sc ii}, and we see that the results are in good agreement.

\subsection{Comparison with normal metal-poor stars of Paper\,{\sc ii}}
\label{comparison_pop2}

To place HE~0338$-$3945 in context, it is of interest to compare this star with ``normal'' Population\,{\sc ii} stars of similar metallicity. We extracted all stars from Paper\,{\sc ii} with $-2.6 < \mathrm{[Fe/H]} < -2.2$. It was required that the stars should have $\mathrm{[Eu/Fe]} < 1$ and $\mathrm{[C/Fe]} < 1$ to be considered as ``normal''. The mean value of [X/Fe] was then computed for the sample where more than five measurements were available. 

\begin{figure}
\begin{center}
\resizebox{\hsize}{!}{\rotatebox{0}{\includegraphics{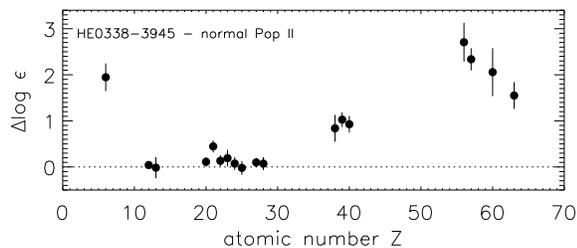}}}
\end{center}
\caption{Difference between abundances for HE~0338$-$3945 and those for normal Population\,{\sc ii} stars of similar metallicity from Paper\,{\sc ii}, derived as described in the text. The vertical bars show the standard deviation among the normal Population\,{\sc ii} stars.}
\label{fig:compare_meanpop2}
\end{figure}

In Fig.\,\ref{fig:compare_meanpop2} we compare the abundances for HE~0338$-$3945 with the described mean abundances.  For reference, the derived mean abundances and standard deviations are provided in Table\,\ref{tab:meanpop2}. In Fig.\,\ref{fig:compare_meanpop2} we clearly see the enhancements of C and the $n$-capture elements relative to the normal stars. For the most part the iron-peak element abundances for HE~0338$-$3945 appear to be normal; however, Sc appears to be significantly enhanced compared to the normal stars. The good agreement found in Sect.~\ref{comparison_paper2} with results for this star from Paper\,{\sc ii} means that this cannot be due to a difference in the analysis, e.g. use of different lines or atomic data.  However, as discussed in Sect.~\ref{results} the Sc abundance may be spurious.  

\begin{table}
\begin{center}
\caption{Comparison of the mean abundances relative to Fe, for normal Population\,{\sc ii} stars of Paper\,{\sc ii} in the metallicity range $-2.6 < \mathrm{[Fe/H]} < -2.2$, with those for HE\,0338-3945. For each element we report the number of measurements $N$ available from Paper\,{\sc ii}, the mean value of [X/Fe], and the standard deviation of [X/Fe] for the sample, the abundance for HE\,0338-3945 $\mathrm{[X/Fe]}_\star$, and the difference $\mathrm{[X/Fe]}_\star - \langle \mathrm{[X/Fe]} \rangle$. }
\label{tab:meanpop2}
\scriptsize
\begin{tabular}{lrrrrr}
\hline
\hline
Element &  $N$ &   $\langle \mathrm{[X/Fe]} \rangle$  & $\sigma(\mathrm{[X/Fe]})$ & $\mathrm{[X/Fe]}_\star$ & $\Delta$[X/Fe]\\
\hline
   C  &  58    & $	  0.18   $ &	  0.30  &   2.13  &    1.95  \\
  Mg  &  58    & $	  0.26   $ &	  0.08  &   0.30  &    0.04  \\
  Al  &  55    & $	 -0.86   $ &	  0.23  & $-0.88$ &  $-0.02$ \\
  Ca  &  58    & $	  0.27   $ &	  0.10  &   0.38  &    0.11  \\
  Sc  &  58    & $	  0.08   $ &	  0.13  &   0.53  &    0.45  \\
  Ti  &  58    & $	  0.24   $ &	  0.12  &   0.37  &    0.13  \\
   V  &  18    & $	  0.00   $ &	  0.18  &   0.19  &    0.19  \\
  Cr  &  58    & $	 -0.19   $ &	  0.14  & $-0.12$ &    0.07  \\
  Mn  &  58    & $	 -0.47   $ &	  0.15  & $-0.49$ &  $-0.02$ \\
  Co  &  53    & $	  0.13   $ &	  0.10  &   0.23  &    0.10 \\
  Ni  &  58    & $	 -0.06   $ &	  0.14  &   0.01  &    0.07 \\
  Sr  &  58    & $	 -0.10   $ &	  0.29  &   0.74  &    0.84 \\
   Y  &  35    & $	 -0.20   $ &	  0.16  &   0.83  &    1.03 \\
  Zr  &  16    & $	  0.27   $ &	  0.18  &   1.20  &    0.93 \\
  Ba  &  55    & $	 -0.30   $ &	  0.42  &   2.41  &    2.71 \\
  La  &   6    & $	 -0.06   $ &	  0.24  &   2.28  &    2.34 \\
  Nd  &  10    & $	  0.24   $ &	  0.52  &   2.30  &    2.06 \\
  Eu  &  21    & $	  0.39   $ &	  0.29  &   1.94  &    1.55 \\
\hline
\end{tabular}
\end{center}
\end{table} 

\subsection{Comparisons with the star HE~2148$-$1247}
\label{comparison_cohen}

Soon after examining our results for HE~0338$-$3945, it became apparent that this star was very similar to the star HE~2148$-$1247 studied by Cohen~et~al.~(\cite{cohen2003}). The Keck/HIRES spectrum of this object used by Cohen~et~al.\ was kindly provided to us by Judith Cohen. Order merging and continuum placement was performed by us; all other reduction steps had been performed. Thus, using our automated spectrum analysis technique we were easily able to perform an analysis of this spectrum homogeneous with our analysis of HE~0338$-$3945. Note, as the spectral coverage of the HE~2148$-$1247 spectrum was not as extensive as for our HE~0338$-$3945 spectrum, we produced a spectrum of HE~0338$-$3945 with identical wavelength coverage to the HE~2148$-$1247 spectrum, which we refer to as the ``cut'' spectrum of HE~0338$-$3945. Thus, an analysis of the cut spectrum permits a more strictly homogeneous analysis as exactly the same spectral lines are employed. 

\begin{figure}
\begin{center}
\resizebox{\hsize}{!}{\rotatebox{0}{\includegraphics{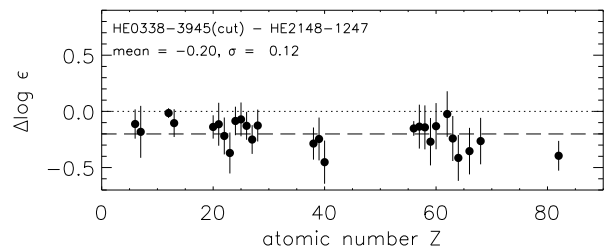}}}
\end{center}
\caption{Difference between abundances for HE~0338$-$3945 (cut spectrum) and HE~2148$-$1247.  The dashed line shows the mean difference.}
\label{fig:compare_he2148}
\end{figure}

For our analysis of HE~2148$-$1247, we adopted the effective temperature from Cohen~et~al., $T_\mathrm{eff} = 6380$\,K. The abundance results from our analysis of HE~2148$-$1247 and the cut spectrum of HE~0338$-$3945 are compared in Fig.\,\ref{fig:compare_he2148}. The abundance patterns of the two stars are seen to be very similar. An offset of $-0.2$~dex between the abundances is seen, but this could easily be due to errors in $T_\mathrm{eff}$ which were not derived consistently. The scatter is consistent with the uncertainties. From this we conclude that the stars have identical abundance patterns for the elements considered, within the errors in the analysis.

\subsection{Comparisons with other similar stars}
\label{comparion_stars}

\begin{table*}
\setlength{\tabcolsep}{0.2mm}
\begin{scriptsize}
\begin{center}
\caption{
Data for stars of different classes. For stars r-I stars and stars not $n$-enhanced, see Paper\,{\sc ii}, Burris et~al.~(\cite{burris2000}), and Fulbright~(\cite{fulbright2000}), although data are presented for some outliers of interest from Burris et al. When several investigations exist, either the most reliable source or the mean of recent data has been taken. The mean of abundance ratios $\mathrm{[X/Y]}$, Y$\ne$Fe, was calculated from the mean of the abundances relative to Fe, and not from individual measurements of $\mathrm{[X/Y]}$. (:) indicates an insecure measurement, and such cases are not considered when calculating the means. References:
(1) This work,
(2) Aoki et~al.~(\cite{aoki2001}),
(3) Aoki et~al.~(\cite{aoki2002}),
(4) Barbuy et~al.~(\cite{barbuy2005}),
(5) Paper\,{\sc ii}
(6) Unpublished preliminary result obtained in Paper\,{\sc ii},
(7) Burris et~al.~(\cite{burris2000}),
(8) Cohen et~al.~(\cite{cohen2003}),
(9) Hill et~al.~(\cite{hill2000}),
(10) Hill et~al.~(\cite{hill2002}) and Plez et~al.~(\cite{plez2004}),
(11) Honda et~al.~(\cite{honda2004a}),
(12) Ivans et~al.~(\cite{ivans2005}),
(13) Johnson \& Bolte~(\cite{johnson2002}),
(14) Johnson \& Bolte~(\cite{johnson2004}),
(15) Jonsell et~al.~(\cite{jonsell2005}),
(16) Lucatello et~al.~(\cite{lucatello2003}),
(17) Preston \& Sneden~(\cite{preston2001}),
(18) Sivarani et~al.~(\cite{sivarani2004}),
(19) Sneden et~al.~(\cite{sneden2003})
(20) Van Eck et~al.~(\cite{van_eck2003}),
(21) Zacs et~al.~(\cite{zacs1998}),
(22) mean of presented data,
(23) mean of (4) and (17).}
\label{Tstars}
\begin{tabular*}{18.0cm}{@{\extracolsep\fill}lllrrrrrrrrrrrrrrr}
\hline
\hline 
\multicolumn{18}{l}{r+s stars} \\ [0.5ex]                                                                                                                                                                 
Name        & T$_\mathrm{eff}$ & $\log g$&[Fe/H]&[C/Fe]&[Sr/Fe]&[Y/Fe]&[Zr/Fe]&[Ba/Fe]&[La/Fe]&[Ce/Fe]&[Eu/Fe]&[Pb/Fe]&[Ba/Eu]&[La/Eu]&[Ce/Eu]&[Pb/Ba]&Ref. \\  										     
\hline
CS~22183-015 & 5200 & 2.5  & $-$3.12 & 2.2: &      & 0.45 & 0.62 & 2.09 & 1.59 & 1.55 & 1.39 & 3.17 & 0.70 & 0.20 & 0.16 & 1.08 & (13)  \\ 
CS~22898-027 &      &      & $-$2.20 & 2.08 & 0.76 & 0.84 & 1.20 & 2.25 & 2.21 & 2.13 & 1.91 & 2.84 & 0.34 & 0.30 & 0.22 & 0.59 & (22) \\ 
            & 6250 & 3.7  & $-$2.25 & 2.2  & 0.92 & 0.73 & 1.01 & 2.23 & 2.13 & 2.13 & 1.88 & 2.84 & 0.35 & 0.25 & 0.25 & 0.61 & (3)  \\ 
            & 6300 & 4.0  & $-$2.15 & 1.95 & 0.60 & 0.95 & 1.39 & 2.27 & 2.28 &      & 1.94 &      & 0.33 & 0.34 &      &      & (17)  \\ 
CS~22948-027 &      &      & $-$2.52 & 2.27 & 0.90 & 1.00 &      & 1.97 & 2.32 & 2.20 & 1.88 & 2.72 & 0.09 & 0.44 & 0.32 & 0.75 & (23)  \\ 
            & 4800 & 1.8  & $-$2.47 & 2.43 & 0.90 & 1.00 &      & 2.26 & 2.32 & 2.20 & 1.88 & 2.72 & 0.38 & 0.44 & 0.32 & 0.46 & (4) \\ 
            & 4600 & 0.8  & $-$2.57 & 2.10 &      &      &      & 1.67 &      &      &      &      &      &      &      &      & (17)  \\ 
            & 4800 & 1.80 & $-$2.46 & 1.8  & 0.90 & 1.00 &      & 2.26 & 2.32 & 2.20 & 2.10 &      & 0.16 & 0.22 & 0.10 &      & (9)  \\ 
CS~29497-030 &      &      & $-$2.64 & 2.43 & 1.09 & 0.84 & 1.42 & 2.25 & 2.16 & 2.12 & 1.72 & 3.60 & 0.53 & 0.44 & 0.40 & 1.35 & (22)  \\ 
            & 7000 & 4.1  & $-$2.57 & 2.47 & 1.34 & 0.97 & 1.40 & 2.32 & 2.22 & 2.10 & 1.99 & 3.65 & 0.33 & 0.23 & 0.11 & 1.33 & (12)  \\ 
            & 6650 & 3.5  & $-$2.70 & 2.38 & 0.84 & 0.71 & 1.43 & 2.17 & 2.10 & 2.14 & 1.44 & 3.55 & 0.73 & 0.66 & 0.70 & 1.38 & (18)  \\ 
CS~29497-034 & 4800 & 1.8  & $-$2.90 & 2.63 & 1.00 & 1.10 &      & 2.03 & 2.12 & 1.95 & 1.80 & 2.95 & 0.23 & 0.32 & 0.15 & 0.92 & (4)  \\ 
            & 4800 & 1.80 & $-$2.91 & 2.3  & 1.00 & 1.10 &      & 2.03 & 2.12 & 1.95 & 2.25 &      &$-$0.22 &$-$0.13 &$-$0.30 &      & (9)  \\ 
CS~29526-110 & 6500 & 3.2  & $-$2.38 & 2.2  & 0.88 &      & 1.11 & 2.11 & 1.69 & 2.01:& 1.73 & 3.30 & 0.38 &$-$0.04 & 0.28 & 1.19 & (3)  \\ 
CS~31062-012 & 6250 & 4.5  & $-$2.55 & 2.1  & 0.30 & 0.59 &      & 1.98 & 2.02 & 2.12 & 1.62 & 2.40 & 0.36 & 0.40 & 0.50 & 0.42 & (3) \\ 
CS~31062-050 &      &      & $-$2.37 & 2.00 & 0.91 & 0.48 & 0.94 & 2.55 & 2.28 & 2.02 & 1.82 & 2.86 & 0.73 & 0.46 & 0.20 & 0.31 & (22)  \\ 
            & 5500 & 2.70 & $-$2.42 & 2.00 &      & 0.48 & 0.85 & 2.80 & 2.12 & 2.02 & 1.79 & 2.81 & 1.01 & 0.33 & 0.23 & 0.01 & (14)  \\ 
            & 5600 & 3.0  & $-$2.32 & 2.00 & 0.91 &      & 1.02 & 2.30 & 2.44 & 2.10:& 1.84 & 2.90 & 0.46 & 0.60 & 0.26 & 0.60 & (3)  \\ 
HE~0024$-$2523 & 6625 & 4.3  & $-$2.71 & 2.6  & 0.34 &$<$0.91 &$<$1.22 & 1.46 & 1.80 &      &$<$1.10 & 3.30 &$>$0.36 &$>$0.70 &      & 1.84 & (16) \\ 
HE~0131$-$3953 & 5928 & 3.83 & $-$2.71 & 2.45 & 0.46 &      &      & 2.20 & 1.94 & 1.93 & 1.62 &      & 0.58 & 0.32 & 0.31 &      & (5)  \\ 
HE~0338$-$3945 & 6160 & 4.13 & $-$2.42 & 2.13 & 0.74 & 0.83 & 1.20 & 2.41 & 2.28 & 2.16 & 1.94 & 3.10 & 0.47 & 0.34 & 0.22 & 0.69 & (1) \\ 
            & 6162 & 4.09 & $-$2.41 & 2.07 & 0.73 & 0.73 &      & 2.41 & 2.26 & 2.21 & 1.89 &      & 0.52 & 0.37 & 0.32 &      & (5)  \\ 
HE~1046$-$1352 & 5540 & 3.0:  & $-$2.6:  & 2.2:  & 0.8:  & 0.5:  &      & 2.1:  & 1.8:  & 1.8:  & 1.4:  &      & 0.7:  & 0.4:  & 0.4:  &      & (6)  \\ 
HE~1105$+$0027 & 6132 & 3.45 & $-$2.42 & 2.00 & 0.73 & 0.75 &      & 2.45 & 2.10 &      & 1.81 &      & 0.64 & 0.29 &      &      & (5)  \\ 
HE~1405$-$0822 & 5400 & 2.0:  & $-$2.1:  & 1.7:  &      & 0.6:  & 0.8:  & 1.8:  & 1.2:  & 1.3:  & 1.1:  &      & 0.7:  & 0.1:  & 0.2:  &      & (6) \\ 
HE~2148$-$1247 & 6380 & 3.9  & $-$2.28 & 1.91 & 0.76 & 0.83 & 1.47 & 2.36 & 2.38 & 2.28 & 1.98 & 3.12 & 0.38 & 0.40 & 0.30 & 0.76 & (8)  \\ 
LP~625$-$44    & 5500 & 2.8  & $-$2.71 & 2.1  & 1.15 & 0.99 & 1.34 & 2.74 & 2.46 & 2.27 & 1.97 & 2.55 & 0.77 & 0.49 & 0.30 & $-$0.19 & (2)  \\ 
LP~706$-$7     & 6000 & 3.8  & $-$2.74 & 2.15 & 0.15 & 0.25 &$<$1.16 & 2.01 & 1.81 & 1.86 & 1.40 & 2.28 & 0.61 & 0.41 & 0.46 & 0.27 & (2) \\ 
\hline
\hline
\multicolumn{18}{l}{s stars} \\  [0.5ex]                                                                                                                                                                
Name        & T$_\mathrm{eff}$ & $\log g$&[Fe/H] &[C/Fe]&[Sr/Fe]&[Y/Fe]&[Zr/Fe]&[Ba/Fe]&[La/Fe]&[Ce/Fe]&[Eu/Fe]&[Pb/Fe]&[Ba/Eu]&[La/Eu]&[Ce/Eu]&[Pb/Ba]&Ref. \\									     
\hline
CS~22880-074 &      &      & $-$1.85 & 1.41 & 0.27 & 0.16 & 0.73 & 1.33 & 1.16 & 1.22 & 0.53 & 1.9  & 0.80 & 0.63 & 0.69 & 0.57 & (22)  \\ 
              & 5850 & 3.8  & $-$1.93 & 1.3  & 0.39 & 0.16 &      & 1.31 & 1.07 & 1.22 & 0.5  & 1.9  & 0.81 & 0.57 & 0.72 & 0.59 & (3)   \\ 
              & 6050 & 4.0  & $-$1.76 & 1.51 & 0.14 & 0.6: & 0.73 & 1.34 & 1.24 &      & 0.55 &      & 0.79 & 0.69 &      &      & (17)  \\ 
CS~22881-036 & 6200 & 4.0  & $-$2.06 & 1.96 & 0.59 & 1.01 & 0.95 & 1.93 & 1.59 &      & 1.00 &      & 0.93 & 0.59 &      &      & (17)  \\ 
CS~22942-019 &      &      & $-$2.66 & 2.0  & 1.4  & 1.58 & 1.69 & 1.71 & 1.53 & 1.54 & 0.79 &$\le$1.6&0.92& 0.74 & 0.75 &$\le-$0.11 & (22)  \\ 
              & 5000 & 2.4  & $-$2.64 & 2.0  & 1.7: & 1.58 & 1.69 & 1.92 & 1.20 & 1.54 & 0.79 &$\le$1.6&1.13& 0.41 & 0.75 &$\le-$0.32 & (3)   \\ 
              & 4900 & 1.8  & $-$2.67 & 2.2: & 1.4  &      &      & 1.50 & 1.85 &      & 0.8: &      & 0.7: & 1.05:&      &      & (17)  \\ 
CS~30301-015 & 4750 & 0.8  & $-$2.64 & 1.6  & 0.3: & 0.29 &      & 1.45 & 0.84 & 1.16 & 0.2:& 1.7  & 1.25:& 0.64:& 0.96:& 0.25 & (3)   \\ 
HD\,196944    &      &      & $-$2.33 & 1.31 & 0.84 & 0.57 & 0.63 & 1.26 & 1.13 & 1.20 & 0.17 & 2.00 & 1.09 & 0.96 & 1.03 & 0.74 & (22)  \\ 
            & 5250 & 1.7  & $-$2.23 &      &      &      &      & 1.14 &      &      &      &      &      &      &      &      & (15)  \\ 
            & 5250 & 1.7  & $-$2.4  &      &      &      & 0.6  &      & 1.00 & 1.10 &      & 2.10 &      &      &      &      & (20)  \\ 
            & 5250 & 1.8  & $-$2.25 & 1.2  & 0.84 & 0.56 & 0.66 & 1.10 & 0.91 & 1.01 & 0.17 & 1.90 & 0.93 & 0.74 & 0.84 & 0.80 & (3)  \\ 
            & 5250 & 1.7  & $-$2.45 & 1.42 &      & 0.58 &      & 1.56 & 1.49 & 1.50 &      &      &      &      &      &      & (21)  \\ 
HE~0202$-$2204 & 5280 & 1.65 & $-$1.98 & 1.16 & 0.57 & 0.41 & 0.47 & 1.41 & 1.36 & 1.30 & 0.49 &      & 0.92 & 0.87 & 0.81 &      & (5)  \\ 
HE~1135$+$0139 & 5487 & 1.80 & $-$2.33 & 1.19 & 0.66 & 0.36 & 0.46 & 1.13 & 0.93 & 1.17 & 0.33 &      & 0.80 & 0.60 & 0.84 &      & (5)  \\ 

\hline
\hline
\multicolumn{18}{l}{r-II stars} \\  [0.5ex]                                                                                                                                                                
Name        & T$_\mathrm{eff}$ & $\log g$&[Fe/H]&[C/Fe]&[Sr/Fe]&[Y/Fe]&[Zr/Fe]&[Ba/Fe]&[La/Fe]&[Ce/Fe]&[Eu/Fe]&[Pb/Fe]&[Ba/Eu]&[La/Eu]&[Ce/Eu]&[Pb/Ba]&Ref. \\									     
\hline
CS~22183-031 & 5270 & 2.80 &$-2.93$ & 0.42 & 0.10 & 0.21 &      & 0.38 &      &      & 1.16 &       &         &        &       &        &  (11) \\ 
CS~22892-052 & 4800 & 1.50 &$-$3.1  & 0.88 & 0.61 & 0.44 & 0.82 & 0.99 & 1.09 & 1.02 & 1.64 &$<$0.9  &$-$0.65 &$-$0.55 &$-$0.62&$<-$0.09& (19) \\ 
            & 4884 & 1.81 &$-$2.95 & 1.00 & 0.61 & 0.45 &      & 1.19 & 1.02 &      & 1.54 &      &$-$0.35 &$-$0.52 &      &      & (5) \\ 
CS~29491-069 & 5103 & 2.45 &$-$2.81 & 0.18 & 0.07 & 0.00 &      & 0.34 &      &      & 1.06 &      &$-$0.72 &      &      &      & (5) \\ 
CS~29497-004 & 5013 & 2.23 &$-$2.81 & 0.22 & 0.57 & 0.66 & 0.94 & 1.21 & 1.21 &      & 1.62 &      &$-$0.41 &$-$0.41 &      &      & (5) \\ 
CS~31082-001 & 4825 & 1.5  &$-$2.92 & 0.2  & 0.65 & 0.43 & 0.73 & 1.17 & 1.13 & 1.01 & 1.63 & 0.40 &$-$0.46 &$-$0.50 &$-$0.62 &$-$0.77 & (10) \\ 
            & 4922 & 1.90 &$-$2.78 & 0.22 & 0.53 & 0.56 & 0.89 & 1.18 & 1.18 & 1.06 & 1.66 &      &$-$0.48 &$-$0.48 &$-$0.60 &      & (5) \\ 
HE~0430$-$4901 & 5296 & 3.12 &$-$2.72 & 0.09 &$-$0.01 & 0.02 &      & 0.50 &      &      & 1.16 &      &$-$0.66 &      &      &      & (5) \\ 
HE~0432$-$0923 & 5131 & 2.64 &$-$3.19 & 0.24 & 0.47 & 0.51 & 0.88 & 0.72 &      &      & 1.25 &      &$-$0.53 &      &      &      & (5) \\ 
HE~1127$-$1143 & 5224 & 2.64 &$-$2.73 & 0.54 & 0.24 & 0.22 &      & 0.63 &      &      & 1.08 &      &$-$0.45 &      &      &      & (5) \\ 
HE~1219$-$0312 & 5140 & 2.40 &$-$2.81 &$-$0.08 & 0.20 & 0.29 & 0.65 & 0.51 & 0.91 &      & 1.41 &      &$-$0.90 &$-$0.50 &      &      & (5) \\ 
HE~2224+0143 & 5198 & 2.66 &$-$2.58 & 0.35 & 0.23 & 0.13 & 0.58 & 0.59 & 0.65 &      & 1.05 &      &$-$0.46 &$-$0.40 &      &      & (5) \\ 
HE~2327$-$5642 & 5048 & 2.22 &$-$2.95 & 0.43 & 0.31 & 0.12 &      & 0.66 & 0.67 &      & 1.22 &      &$-$0.56 &$-$0.55 &      &      & (5) \\ 
\hline
\hline
\multicolumn{18}{l}{Outlier ``normal'' stars} \\ [0.5ex]                                                                                                                                                                 
Name        & T$_\mathrm{eff}$ & $\log g$&[Fe/H]&[C/Fe]&[Sr/Fe]&[Y/Fe]&[Zr/Fe]&[Ba/Fe]&[La/Fe]&[Ce/Fe]&[Eu/Fe]&[Pb/Fe]&[Ba/Eu]&[La/Eu]&[Ce/Eu]&[Pb/Ba]&Ref. \\									     
\hline
HD\,13979     & 5075 & 1.90 & $-$2.26 &      &$-$0.07 &$-$0.63 &$-$0.34 &$-$0.50 &$-$0.10 &      &$-$0.38 &      &$-$0.12 & 0.28 &      &      & (7) \\ 
HD\,25532     & 5300 & 1.90 & $-$1.46 &      & 0.14 &$-$0.28 & 0.14 & 0.17 &$-$0.07 &      & 0.10 &      & 0.07 &$-$0.17 &      &      & (7) \\ 
HD\,105546    & 5300 & 2.50 & $-$1.27 &      & 0.45 & 0.13 & 0.45 & 0.42 & 0.05 &      & 0.32 &      & 0.10 &$-$0.27 &      &      & (7) \\ 
HD\,166161    & 5150 & 2.20 & $-$1.30 &      & 0.20 & 0.31 & 0.43 & 0.53 & 0.36 &      & 0.10 &      & 0.43 & 0.26 &      &      & (7) \\ 
HD\,218857    & 5125 & 2.40 & $-$1.86 &      & 0.01 &$-$0.17 & 0.00 & 0.03 &$-$0.36 &      &$-$0.23 &      & 0.26 &$-$0.13 &      &      & (7) \\ 
BD$+$11$^\circ$2998  & 5425 & 2.30 & $-$1.17 &      &$-$0.20 &$-$0.23 & 0.17 & 0.14 &$-$0.17 &      & 0.06 &      & 0.08 &$-$0.23 &      &      & (7) \\ 
BD$+$17$^\circ$3248  & 5250 & 2.30 & $-$2.02 &      & 0.55 & 0.06 & 0.50 & 0.97 & 0.60 &      & 0.96 &      & 0.01 &$-$0.36 &      &      & (7) \\ 
\hline
\hline
\end{tabular*}
\end{center}
\end{scriptsize}
\end{table*}

In addition to HE~0338$-$3945 and HE~2148$-$1247 there are several other stars that have recently been found to show both enhanced $r$- and $s$-elements. In total 17 such stars are known to us, and these are listed in Table\,\ref{Tstars}. 

We have investigated the similarities within the $r$- and $s$-element enhanced group, and the differences relative to other metal-poor stars. The $r$- and $s$-enhanced stars were compared with other stars of [Fe/H]$<-1.0$ investigated in Paper\,{\sc ii}, by Burris et~al.~(\cite{burris2000}), and Fulbright~(\cite{fulbright2000}), and some other groups listed in Table\,\ref{Tstars}. When several sources were available, data were adopted from the source considered most reliable or the mean of recent data was taken.

The $r$- and $s$-enhanced stars HE~1046$-$1352 and HE~1405$-$0822 were analysed in the work of Paper\,{\sc ii}, but unpublished due to the significant pollution of the spectrum by molecular carbon features making the analysis method uncertain.   The presented data for these stars must therefore be regarded as uncertain and preliminary.  However, higher quality data of the star HE~1405$-$0822 is considered in the HERES paper Sivarani et~al.~(in preparation). It must also be observed that the error bars were quite large for some stars in several investigations, and that errors in stellar parameters may be the reason for stars with seemingly peculiar abundance patterns.

\begin{figure}
\centering
\resizebox{\hsize}{!}{\rotatebox{-90}{\includegraphics*{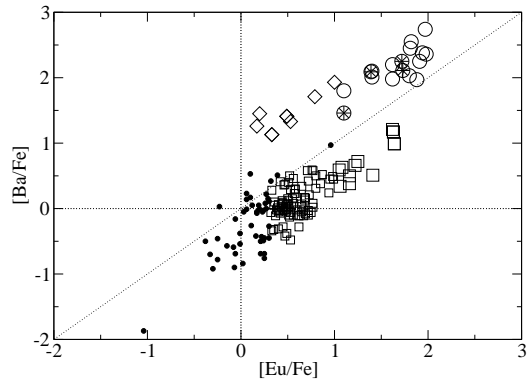}}}
\caption[]{\label{Feufebafe}
A plot comparing the abundance of the $s$-element Ba to the $r$- element Eu for stars of different classes (see definitions in Table\,\ref{Tclasses}). The r+s stars, enhanced in both $r$- and $s$-elements, are shown as open circles, and the r+s stars also known to be enhanced in Pb are shown as circles with asterisks inside.  The stars enriched only in $r$-elements are marked as small (r-I) and large (r-II) open squares, and stars with only $s$-enhancement are shown as diamonds. Stars not enhanced in either $r$- or $s$-elements, in this article called normal stars, are marked by small filled circles. The dotted lines mark one-to-one relations. Data are taken from Table\,\ref{Tstars} and for normal and r-I stars from Paper\,{\sc ii}, Burris et~al.~(\cite{burris2000}), and Fulbright~(\cite{fulbright2000}).}
\end{figure}

The plot in Fig.\,\ref{Feufebafe} shows a clear separation of different kinds of stars. The criteria of the classes in Table\,\ref{Tclasses} in Sect.\,\ref{discussion} were defined to achieve this. The stars were classified as r-I, r-II, $s$-element enhanced, r+s, i.e.\ both $r$- and $s$-element enhanced, and/or Pb enhanced. Stars with no significant enhancement, and thus not falling under any of these criteria, are in this paper called ``normal'' metal-poor stars. 

The r+s stars are rich in both $r$- and $s$-elements, residing in the upper right corner of Fig.\,\ref{Feufebafe}. All of these stars have $\mathrm{[Ba/Eu]} > 0.0$. This is also true for the Pb stars with measurements of both Ba and Eu found in the literature. These Pb stars are also then classified as r+s stars. The $s$-enhanced stars also have $\mathrm{[Ba/Eu]} > 0.0$, and may be located along the same sequence in the figure as the r+s stars. The r-II and r-I stellar classes both have $\mathrm{[Ba/Eu]} < 0.0$. The r-II stars have in general a typically lower content of $\mathrm{[Eu/Fe]}$ than r+s stars, and in Fig.\,\ref{Feufebafe} seem to separate clearly from the r+s stars.  The gap in between the groups might be due to poor statistics; this separation was also noted by Cohen et~al.~(\cite{cohen2003}) (however, with some overlap). The normal stars, i.e. stars with no $n$-capture element enhancement, have predominantly $\mathrm{[Ba/Eu]} < 0.0$, although positive ratios may also be found. The same pattern was also found by Cohen et~al.~(\cite{cohen2003}, Fig.\,13) and Johnson \& Bolte (\cite{johnson2004}, Fig.\,5), although the nomenclature is different in these studies (see Sect.\,\ref{conclusions}).

\begin{figure}
\begin{center}
\resizebox{\hsize}{!}{\rotatebox{-90}{\includegraphics*{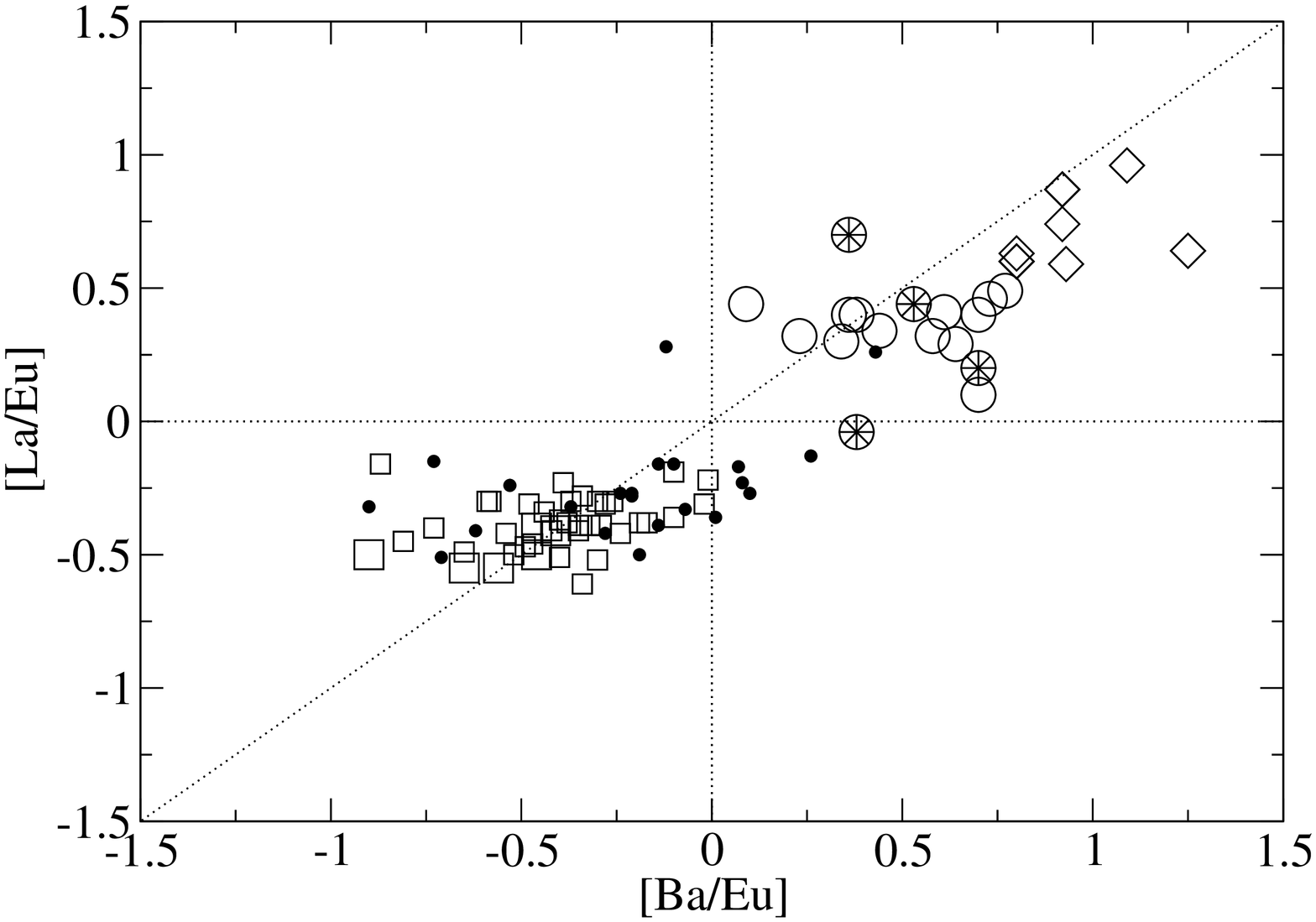}}}
\resizebox{\hsize}{!}{\rotatebox{-90}{\includegraphics*{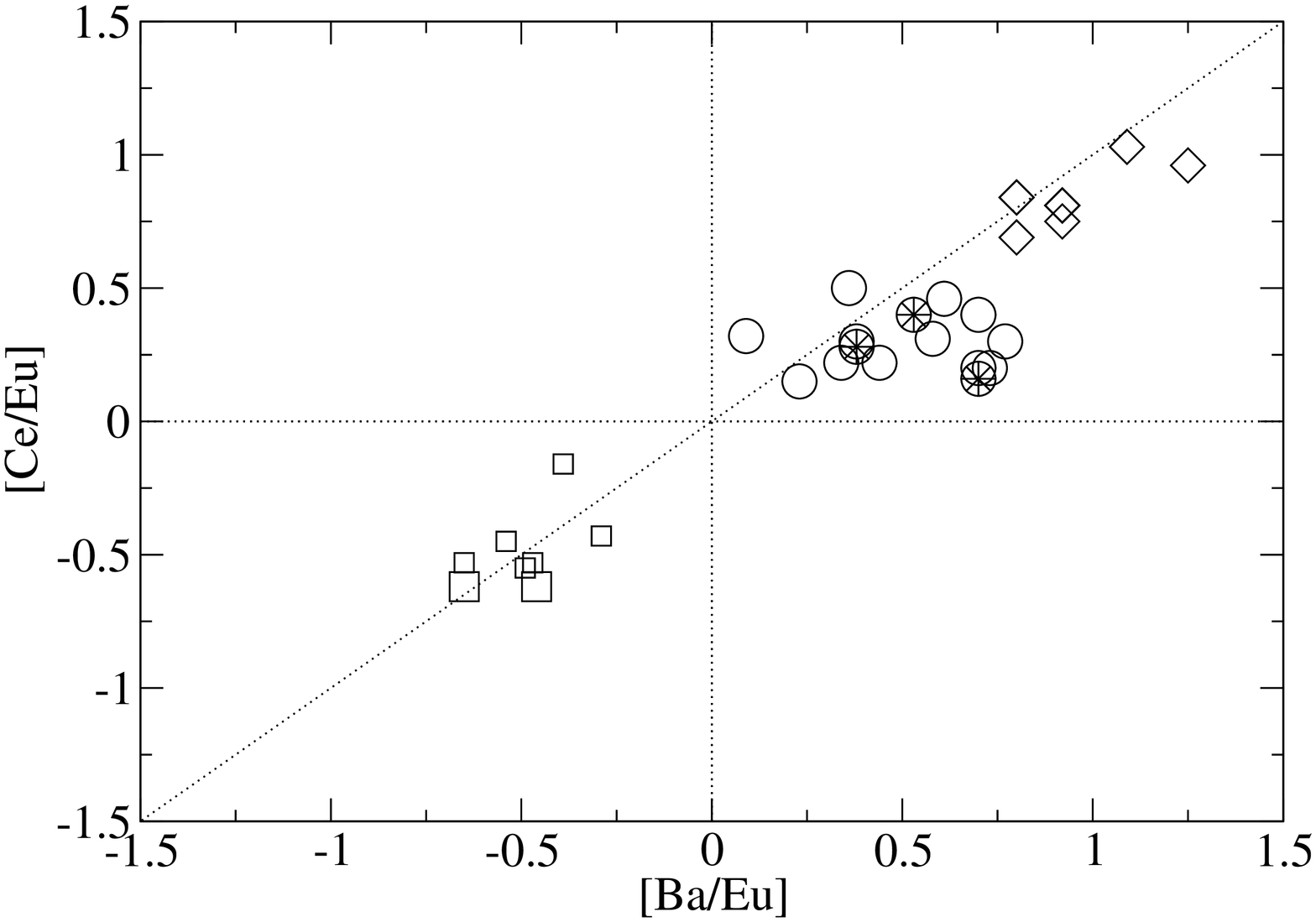}}}
\resizebox{\hsize}{!}{\rotatebox{-90}{\includegraphics*{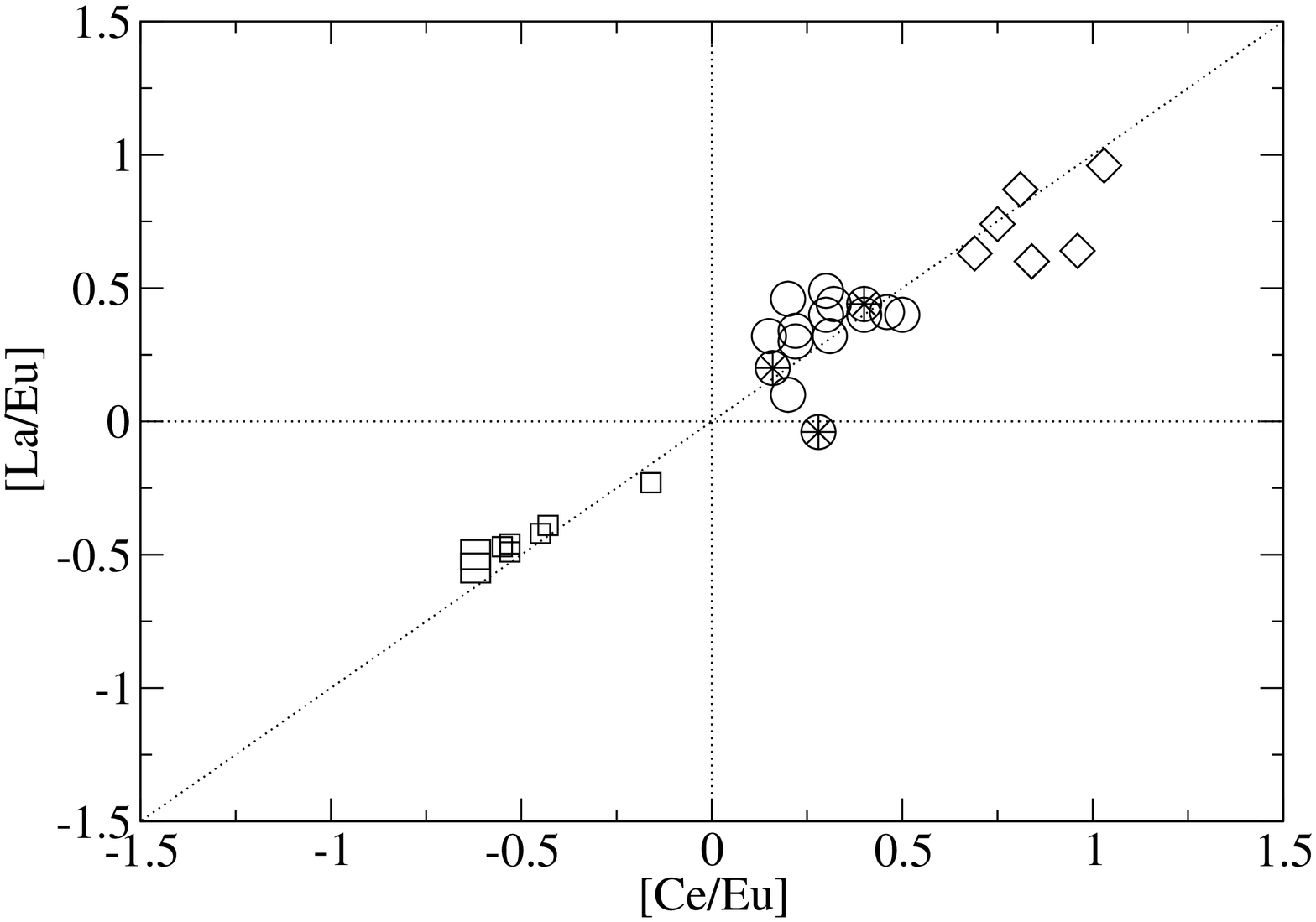}}}
\end{center}
\caption[]{\label{Fbaeulaeuceeu}
Plots showing the ratio of the different heavy $n$-capture $s$-elements (Ba, La, Ce) compared to the $r$-element Eu for stars of different classes. Symbols and data as in Fig.\,\ref{Feufebafe}.}
\end{figure}

The plots in Fig.\,\ref{Fbaeulaeuceeu} show the ratio of the different heavy $n$-capture $s$-elements ($hs$ = Ba, La, Ce) compared to the $r$-element Eu for stars of different classes. As is also seen in Fig.\,\ref{Feufebafe} there is a clear separation between on one hand r+s and s stars, and on the other hand r-II, r-I and normal stars. The different classes of stars seem to follow a single well-defined sequence in the [La/Eu] vs.\ [Ce/Eu] plot, but less so in the other abundance diagrams. $\mathrm{[Ba/Eu]}$ shows the greatest variation amplitude in the diagrams, while La and Ce seem vary less and follow each other.  According to Paper\,{\sc ii} this may well be an observational effect; as seen in Table\,\ref{Terrors} the abundances of Ba and of the other elements react differently to changes in effective temperature and gravity which is essentially a consequence of the significance of the strong damping wings of the dominating \ion{Ba}{ii} line $\lambda 4554$.  Therefore, stellar-parameter errors cancel for the ratios $\mathrm{[La/Eu]}$ and $\mathrm{[Ce/Eu]}$, but not for $\mathrm{[Ba/Eu]}$. The Pb enhanced r+s star CS~29526-110  has a rather low abundance of La (Aoki et~al.~\cite{aoki2002}), which makes the ratio $\mathrm{[La/Eu]}$ negative. This is astonishing since the other $hs$-elements (Ba, Ce) are considerably enhanced. The authors note that the effective temperature and gravity are rather uncertain for this star. 

There are some outliers among the normal stars in the upper panel of Fig.\,\ref{Fbaeulaeuceeu}.  The parameters of these stars may be viewed in Table\,\ref{Tstars}. These stars reside in or quite near the same region as the r+s stars, but are neither $r$- nor $s$-enhanced. The star BD+17$^\circ$3248 is rather close to being regarded as an r+s star, as it has abundances $\mathrm{[Ba/Fe]}=0.97$, $\mathrm{[Eu/Fe]}=0.96$, and $\mathrm{[Ba/Eu]}=0.01$.  However, in Fig.\,\ref{Feufebafe} it lies closer to the normal stars.  We note that Cowan~et~al.~\cite{cowan2002} find a much lower Ba abundance for this star, which would place it firmly as a r-I star.

\begin{figure}
\centering
\resizebox{\hsize}{!}{\includegraphics*{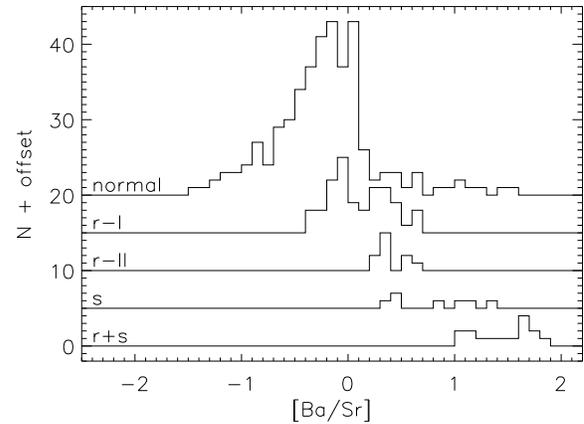}}
\caption[]{\label{Fsrbahist}
Histograms showing the distribution of different classes of stars when comparing the heavy $n$-capture $s$-element Ba to the light $n$-capture $s$-element Sr. Data as in Fig.\,\ref{Feufebafe}.}
\end{figure}

The histograms in Fig.\,\ref{Fsrbahist} display the abundance of the heavy $s$-element Ba compared to the light $s$-element ($ls$) Sr for the different stellar classes. In spite of the small number statistics, there are clear separations in mean ratios for different classes. Normal stars have a slightly negative mean $\mathrm{[Ba/Sr]}$, although the spread is wide. The mean r-I star has slightly more Ba than Sr, and for r-II stars Ba dominates.  The s stars in the diagram are very few, but tend to have even higher $\mathrm{[Ba/Sr]}$ than the r-II stars. The r+s stars are, however, the most overabundant in Ba relative to Sr of all groups, with a mean of $\mathrm{[Ba/Sr]} \sim 1.6$.

We note from Table\,\ref{Tstars} that there may also be a difference in the metallicity distribution of the stars within these groups.  The r-II stars are centred on a metallicity of $\mathrm{[Fe/H]} \sim -2.84$ with a scatter on the order of 0.16~dex (see also Paper\,{\sc ii}).  The r-I stars are found across the entire range of metallicities investigated in Paper\,{\sc ii}, i.e.\ $-3.4 < \mathrm{[Fe/H]} < -1.5$.  The r+s stars from Table\,\ref{Tstars} are centred on $\mathrm{[Fe/H]} \sim -2.55$ with a scatter on the order of 0.26~dex.  Thus, the r+s stars seem on average to have a higher metallicity and a wider range of metallicity than the r-II stars. 

\section{Discussion}
\label{discussion}

In this section we will discuss both the results for HE~0338$-$3945, and the r+s stars in general.  We first summarise the abundances of HE~0338$-$3945 and then briefly discuss the classification of the r+s stars, in the light of our survey of stars with $n$-capture enhancement in Sect.~\ref{comparion_stars}.  Finally, we present a detailed discussion of the possible formation scenarios for the r+s stars, based on both our results for the abundance pattern of HE~0338$-$3945 and our survey of r+s stars from the literature.

\subsection{The abundances of HE~0338$-$3945}
\label{dicussionabundances}

Abundances were derived for 33 elements and we estimated upper limits for an additional 6 elements for the star HE~0338$-$3945. We have confirmed the high content of the $r$-elements Eu, Gd, Dy, Er, and Tm  with a mean of $\mathrm{[r/Fe]} \sim 2.10$ ($\mathrm{[Eu/Fe]} \sim 1.94$). The upper limits on Tb, Ho, Lu support an overall high $r$-element abundance, although the Ag upper limit is surprisingly low and not consistent with a scaled solar $r$-process distribution as normalised to Eu. This could, however, be an effect of over-ionisation of Ag I. We could detect neither Th nor U, as the spectral lines of these elements are weak and reside in heavily blended regions; only rather high upper limits were obtained.

Although the star is also enhanced in $s$-elements, there is a pronounced difference among the $ls$ and the $hs$ nuclei. The mean of the $ls$ overabundance is $\mathrm{[Sr, Y, Zr/Fe]} = 0.92$, and the $hs$ $\mathrm{[Ba, La, Ce/Fe]} = 2.27$. Obviously, the light $s$-elements do not fit the solar $s$-element distribution of $s$-elements when normalised to Ba. However, they do fit the $r$-element distribution as normalised to Eu.  This behaviour has been noted in other stars as well, such as HE~2148$-$1247 (Cohen et~al.~\cite{cohen2003}). The abundance of Pb is higher than expected from scaled solar $s$-element abundances as normalised to Ba, by about 1 dex. The elements that for the Sun were produced in significant amounts by both the $r$- and $s$-processes (e.g.\ Nd, Sm and Yb) are not unexpectedly overabundant compared to both the $r$- and the $s$-element solar distributions.

The overabundance of C is confirmed, and we have derived an isotopic ratio of $^{12}$C/$^{13}$C$\sim 10$. Also N and O are overabundant in the star.  Na to Ni seem normal for matter contaminated by SNe of type {\sc ii}, the exception being Sc which is significantly enhanced compared to normal Population\,{\sc ii} stars (but see Sect.~\ref{results}).  The abundance of Cu is low relative to iron, and comparable with results for normal stars of similar metallicity (Mishenina et~al.~\cite{mishenina2002}, Bihain et~al.~\cite{bihain2004}).

The NLTE and 3D effects discussed in Sect.\,\ref{results} may in many cases be severe. We lack estimates of corrections for most elements, and, moreover, the estimates we have are often quite uncertain. It is thus risky to draw conclusions about nucleosynthesis sites by comparing abundances of different elements. One may hope that, e.g., the net effect of NLTE and 3D could well be of the same order of magnitude for the $ls$-element Sr and the $hs$-element Ba, but the different character of the lines used may speak against such an optimistic view. Comparisons between different groups of elements, when carried out differentially relative to other stars with similar temperatures and gravities and overall metallicities may, however, be discussed with some confidence, in particular when we have some guidance from theoretical estimates from NLTE and 3D calculations.
Unfortunately, 3D corrections for Eu are unavailable at present, and these effects could perhaps be of some importance.

\subsection{\label{r+s}The class r+s stars}

With the background of Sect.\,\ref{comparion_stars} we shall here discuss the classification of r+s stars. We propose to use $\mathrm{[Eu/Fe]}$ as a measure for the $r$-elements, and $\mathrm{[Ba/Fe]}$ for the $s$-elements. This choice of classification criteria, depending on single $n$-capture element abundances instead of e.g.\ all $hs$-elements, is motivated by simplicity and practicality.  It could be of value to base a classification system on many criteria to suppress effects of errors and individual abundance fluctuations, but in practice both Eu and Ba have strong lines which are easy to observe in metal-poor stars and moreover, ambiguities are minimised. Also, this simple classification could be transformed to a purely empirical system, independent of uncertainties in the abundance analysis. 

\begin{table}
\begin{scriptsize}
\begin{center}
\caption{\label{Tclasses} Definition of classes for $n$-capture-rich stars. In this article we define stars not falling under any of these classes as ``normal''.}
\begin{tabular*}{8.8cm}{@{\extracolsep\fill}lll}
\hline
\hline
Class & Constraints in [X/Fe] & Other constraints \\ [0.5ex]
\hline
\\
r-I     & $+0.3 \le \mathrm{[Eu/Fe]} \le +1.0$ & $\mathrm{[Ba/Eu]} < 0.0$ \\
\\
r-II    &          $\mathrm{[Eu/Fe]} > +1.0$   & $\mathrm{[Ba/Eu]} < 0.0$ \\
\\
r+s     &          $\mathrm{[Ba/Fe]} > +1.0$   & $\mathrm{[Ba/Eu]} > 0.0$ \\
        &          $\mathrm{[Eu/Fe]} > +1.0$   &                          \\ 
\\
s       &          $\mathrm{[Ba/Fe]} > +1.0$   & $\mathrm{[Ba/Eu]} > 0.0$ \\
        &          $\mathrm{[Eu/Fe]} \le +1.0$ &                          \\
\\
Pb      &          $\mathrm{[Ba/Fe]} \ge +1.0$ & $\mathrm{[Pb/Ba]} \ge +1.0$ \\
\\
\hline
\end{tabular*}
\end{center}
\end{scriptsize}
\end{table}

The criteria chosen for the r+s stars are presented in Table\,\ref{Tclasses}. It is natural to set the limit for $s$-elements as $\mathrm{[Ba/Fe]} > +1.0$, see Fig.\,\ref{Feufebafe}, but the limit for the $r$-elements may be debated. For simplicity we have made a choice analogous to that for the $s$-elements, and require $\mathrm{[Eu/Fe]} > +1.0$, although there is no natural separation between the r+s stars and the s stars in this respect. However, the distribution of the stars in the [Ba/Eu] versus [Ce/Eu] and [La/Eu] versus [Ce/Eu] diagrams (see Fig.\,\ref{Fbaeulaeuceeu}) suggests that there might be a natural separation of these classes of stars.  As seen from the plots of Fig.\,\ref{Fbaeulaeuceeu} it is also natural to set the criteria $\mathrm{[Ba/Eu]} > 0.0$ for the r+s stars. Note that we do not require the r+s stars to be rich in C, although all such stars known to us have $\mathrm{[C/Fe]} \ge +1.7$

The definition of the class r+s only partially overlaps with the class ``r/s stars'' defined by Beers \& Christlieb (\cite{beers2005}). The r/s criteria is set to $0.0 < \mathrm{[Ba/Eu]} < +0.5$, and thus also includes stars enhanced in neither $r$- nor $s$-elements. The r+s class, on the other hand, includes stars with $\mathrm{[Ba/Eu]} > +0.5$.

As seen in Table\,\ref{Tclasses}, we use the same criteria for r-I and r-II stars as Beers \& Christlieb (\cite{beers2005}), but this is not the case for s stars. Stars highly enhanced in $s$-elements, only moderately in $r$-elements, and with $\mathrm{[Ba/Eu]} > 0.0$ are called s stars here. This definition differs from the classification of Beers \& Christlieb, who define s stars as having $\mathrm{[Ba/Fe]} > +1.0$ and $\mathrm{[Ba/Eu]} > +0.5$. 

Lead stars are defined as $\mathrm{[Pb/hs]}=\mathrm{[Pb/Ba, La, Ce]} > +1.0$ by Van Eck et~al.~(\cite{van_eck2003}). According to the same principles as discussed above, we propose to use only Ba as the measure for the $hs$-elements, and thus advocate the classification given in Table\,\ref{Tclasses}. We do not require the Pb stars to also be r+s stars, only that they are enhanced in Ba and Pb, and much more in the latter element. 

\subsection{Scenarios for the formation of r+s stars }
\label{scenarios}

The abundance patterns of stars enriched in both $r$- and $s$-elements may give important clues to the sites of nucleosynthesis for these elements. Many different scenarios have been proposed, most of them invoking two more or less independent processes, e.g.\ with one site for the production of the $r$-elements and one for the $s$-elements. A basic question is whether such scenarios may be retained, in view of the homogeneity of the r+s group, and its separation from the other classes of Population\,{\sc ii} stars. Another factor of significance is the frequency of r+s stars, relative to stars with more normal Population\,{\sc ii} composition. From Paper\,{\sc ii} we estimate that on the order of 1\% of the Population\,{\sc ii} stars are r+s stars, though we note this may be significantly underestimated due to the bias against CH strong stars. Here, these different r+s scenarios will be discussed.

\paragraph{I. Radiative levitation:}
Peculiar abundance patterns may arise due to radiative levitation of elements with many absorption lines, while heavy elements with few lines may sink in the atmosphere. This is seen in some Ap stars, which display enormous enhancements of the rare-earth elements. 

Cohen et~al.~(\cite{cohen2003}) discussed this possibility as an explanation for the r+s stars and dismissed it. The r+s stars do not show the same abundance patterns as the Ap stars. They presumably are not hot enough to produce any significant radiative levitation in their atmospheres.  In particular, they have much deeper convective zones, efficiently mixing and diluting the surface material with the stellar interior.  Note also that there are several r+s stars on the giant branch, where the convective zone contains most of the stellar mass.

\paragraph{II. $r$-rich ISM and self-pollution:}
The star may have been formed out of $r$-enriched material from e.g.\ an early supernova, and later self-enriched its surface with $s$-elements during He shell flashes in the AGB phase (as discussed by Hill et~al.~\cite{hill2000}, Cohen et~al.~\cite{cohen2003}). We note that low-mass metal-poor stars may also be affected by dredge-up at the He flash (Fujimoto et~al.~\cite{fujimoto2000}). However, both HE~0338$-$3945 and HE~2148$-$1247 are located in the turn-off region and presumably have not passed the red-giant phase yet, so in the case of at least these r+s stars it seems possible to dismiss the hypothesis. 

\paragraph{III. Binary system out of $r$-rich ISM and AGB-pollution:}
The star could have been a secondary in a binary system, formed out of $r$-enriched material, and later polluted with C and $s$-elements by mass transfer from its AGB companion (discussed by e.g.\ Hill et~al.~\cite{hill2000}, Cohen et~al.~\cite{cohen2003}, Ivans et~al.~\cite{ivans2005}). We note in passing that metal-poor stars may well produce $s$-elements even without any Fe seed nuclei, according to Siess et~al.~(\cite{siess2002}). 

The mass transfer from a companion AGB star is the scenario generally adopted for the formation of CH stars, known to be single-line spectroscopic binaries (McClure~\cite{mcclure1983, mcclure1984} and McClure \& Woodsworth~\cite{mcclure1990}). It is not clear, however, if all r+s stars also are CH stars, although all 17 r+s stars in Table\,\ref{Tstars} have considerably enhanced carbon abundances, as noted in Sect.\,\ref{r+s}. Many CH stars are defined as s  stars; the s stars are often quite C rich (see Table\,\ref{Tstars}), while the r stars are generally less C rich. A relatively high fraction of the CH stars seem, however, to be r+s stars (e.g. Aoki et~al.~\cite{aoki2002}).

There are severe constraints set on the period and orbit for this mechanism to result in a C and $s$-enriched matter transfer, as the mass transfer should occur in the AGB phase where the $s$-elements are produced and not (only) in the RGB phase (Jorissen \& Boffin~\cite{jorissen1992}).  The transfer across the binary has to be efficient to reach the levels of contamination seen in r+s stars. Also the right amounts of elements should be transferred.  A glance in Table\,\ref{Tstars} shows that the carbon enrichment is remarkably constant among the r+s stars.  The low $^{12}\mathrm{C}/^{13}\mathrm{C}$ ratio of HE~0338$-$3945 suggests that a rather large fraction of the carbon (which presumably was produced by He burning in the primary) has been CNO-processed; N and O abundances also clearly suggest heavy processing but the dominant C relative to N and O puts constraints on this. There are obviously several steps that have to work, from the initially $r$-enriched material from which the star happened to form, via the formation of carbon, the formation of the $s$-elements in the AGB inter-shell nucleosynthesis, the CNO processing of C and the pollution of the surface of the companion with sufficient amounts of matter. 

One may discuss whether this is a realistic scenario, in view of the many r+s stars discovered. In general, if a mechanism with two independent steps is in action, one producing the $r$-enrichment and another the $s$-enrichment, and the two are stochastically independent, one would expect the probability to find a star affected by both mechanisms, $p(r,s)$ to be the product of the probabilities of finding a star subjected to only one of the two, $p(r)$ and $p(s)$, respectively, i.e.\ $p(r,s)=p(r)\cdot p(s)$.  Presently, in the HERES survey there are 8 r-II stars, two s stars and three r+s stars, among the 253 stars without strong CH lines analysed in Paper\,{\sc ii}. The CH-strong rejection criteria may lead to a strong bias, suppressing the number of s stars, and even r+s stars. In order for a two-step mechanism to produce 3 r+s stars, however, a considerable fraction of all Population\,{\sc ii} stars should show s-star characteristics. In fact, we find from elementary statistical arguments that the probability of finding 3 or more r+s stars out of 250 would be only 0.9\% if the fraction of s stars would be 0.1 of all Population\,{\sc ii} stars. Even if the total fraction of s stars were 0.3, the chance to find 3 r+s stars would be only 22\%. One should also note that in Paper\,{\sc ii} the r+s stars can only be detected when they are warm enough such that the CH lines are not so strong as to render the method used uncertain.  Only a small fraction of the r-II stars found in the HERES survey are so warm, which suggests that the ratio of stars with r+s abundances relative to r-II stars is considerably higher than suggested by the HERES results. Thus, the probability that an r+s star is just a star which happened to become $r$-enriched, e.g. due to initial inhomogeneities in the ISM or a later nearby SN, and independently of that had an AGB enrichment from a companion, is negligible and should be dismissed. Still another argument against such a two-step scenario is the chemical homogeneity within the r+s group and with a clear separation from r-II stars and s stars. The scenario is also problematic as it is not commonly accepted that the ISM was sufficiently inhomogeneous (e.g.\ Qian \& Wasserburg 2001) to contain such great $s$- and $r$-element overabundances relative to Fe at [Fe/H]$> -3$. 

A possible modification of this scenario would be that the formation of the binary system is triggered by a supernova which also provides the $r$-elements (discussed by Gallino~et~al.~\cite{gallino2005} and Ivans~et~al.~\cite{ivans2005}).  Vanhala \& Cameron~(\cite{vanhala1998}) have performed simulations of triggered star formation due to shock waves impacting molecular cloud cores, suggesting that weakly evolved cores may fragment during collapse and form low-mass binaries.  Such a scenario is appealing since the $r$- and $s$-enrichments are no longer stochastically independent and thus the high frequency of r+s stars might be explained.  However, it is presently unclear whether this scenario could be so prevalent as to explain the observed frequency.

Cohen et~al.~(\cite{cohen2003}) argue for a $s$-enhancement due to an AGB contamination, although they prefer scenario V below. They point out that several of the stars we now define as r+s stars are binaries (CS~29526-110, HE~2148$-$1247, HE~0024$-$2523, LP~625$-$44), although we cannot confirm this for HE~0338$-$3945, nor for several other r+s stars. Preston \& Sneden (\cite{preston2001}) report on r+s stars not exhibiting radial velocity variations exceeding 0.5 km/s over an 8 year period. This discussion may also be applied to scenarios IV, V and VI.

\paragraph{IV. Triple system with SN- and AGB-pollution:}
The star could have been a tertiary in a triple system, in which the primary exploded as a $r$-element producing supernova, and the other companion evolved into an AGB star dumping $s$-rich material on the least massive star. 

This scenario is also discussed and dismissed by Cohen et~al.~(\cite{cohen2003}) as being not very plausible. The likelihood that a star would survive a close SN explosion, that the secondary would not drift away, and that the secondary could subsequently transfer processed matter onto the star at the right time does not seem very great. If such events do happen, it is not probable that they would be common enough or sufficiently constrained to create so many r+s stars with similar abundance patterns, notably with similar relative amounts of C and of $r$- and $s$-elements.

\paragraph{V. Binary system with AGB- and 1.5 SN-pollution:}
The star could be a secondary in a binary system where the companion first contributed $s$-elements as an AGB star, and later exploded as a ``Type 1.5'' supernova (Zijlstra~\cite{zijlstra2004}, Wanajo et~al.~\cite{wanajo2005}), producing the $r$-elements.  The SNe of type 1.5 are proposed to be more common among metal-poor stars than for Population\,{\sc i}, due to the strong metallicity dependence of the mass loss during the AGB phase leading to a different  initial-final-mass relation for low-metallicity stars. Thus, intermediate-mass metal-poor stars may end up with higher final masses after the AGB phase, and more easily reach the Chandrasekhar mass and explode as a Type 1.5 supernova.

\paragraph{VI. Binary system with AGB- and AIC-pollution:}
The star could be a secondary in a binary system, in which the companion first contributed $s$-elements as an AGB star and then evolved to a white dwarf.  Later, the primary might have undergone an accretion-induced collapse (AIC) in the white dwarf stage, triggered by mass transfer in the other direction across the system. The collapse created a neutrino wind and $r$-elements which were transferred across the binary and contaminated the surface of the secondary star (Qian \& Wasserburg~\cite{qian2003}, Cohen et~al.~\cite{cohen2003}). Cohen et~al. also noted that the system might disrupt during the AIC event, which would explain an apparently single r+s star.

This hypothesis is physically uncertain due to uncertainties in neutrino and neutron-star physics (Qian \& Woosley~\cite{qian1996}), and it is thus uncertain whether it works at all.  It is also unclear how the white dwarf would accrete matter from the secondary star, which in several cases has been found not to be evolved beyond the turn-off phase (HE~0338$-$3945, HE~2148$-$1247). One may question how probable is it that the white dwarf is close enough for mass transfer, in view of the mass-loss it has experienced which increases the distance between the components. 

\paragraph{VII. Binary system with only AGB-pollution:}
The star might be a secondary in a metal-poor binary system with an AGB companion producing $n$-capture elements in a hypothetical high neutron density $s$-process. The relatively high flux of neutrons has been proposed to not only produce ``normal'' $s$-elements such as Ba and La, but Eu which otherwise is mainly assumed to be produced in the $r$-process with much higher neutron densities.

This hypothesis has been discussed by Aoki et~al.~(\cite{aoki2002}), Cohen et~al.~(\cite{cohen2003}) and Johnson \& Bolte~(\cite{johnson2004}). Aoki et~al.\ and Cohen et~al.\ found some support from the calculations by Goriely \& Mowlavi~(\cite{goriely2000}). Cohen et~al.\ found a ratio of $\mathrm{Ba/Eu}\sim 100$ for the star HE~2148$-$1247 (we similarly find $\mathrm{Ba/Eu}\sim 115$ for HE~0338$-$3945), and Goriely \& Mowlavi predicted a surface abundance ratio of 105 for their most metal-poor (Z=0.001) AGB-star model with dredge up from a long sequence of He shell flashes. This was, however, the resulting ratio from a relatively mild enhancement (0.48 dex in Ba, 0.07 in Eu) of material with initially solar Ba/Fe and Eu/Fe ratios. The dredge-up material in the model of Goriely \& Mowlavi has a Ba/Eu ratio of about 500.  Anyhow, Cohen et~al.\ dismissed the scenario since they found the predicted absolute abundances of the produced elements to be quite insufficient to explain the observations. Johnson \& Bolte, on the other hand, found evidence in the $r$-element ratios (Eu/Tb, Eu/Dy, Th/Eu) which seem difficult to explain with any combination of the normal $r$- and $s$-processes. They suggested that another form of $s$-process is at work, and found some support for this in parametrised calculations made by Malaney~(\cite{malaney1987}). In Malaney's Table~II, yields of the $s$-process acting at the unusually high neutron density of $10^{12}$ cm$^{-3}$ are presented, which reduced the Ba/Eu ratios to $150-200$. Adopting a neutron exposure of $\tau_0 =0.05$ Johnson \& Bolte found the predictions of Malaney to agree reasonably well with the observed La/Eu ratio for the r+s star CS~31062-050, and a generally good overall fit of the abundance pattern of that star from La to Hf. They noted, however, that the observations of the $^{151}\mathrm{Eu}/^{153}\mathrm{Eu}$ ratio of this star by Aoki et~al.~(\cite{aoki2002}) seem to indicate that the neutron density was rather around the more normal value of $10^8$.

In order to study the consequences of a high neutron-density s process we set up a small program to calculate $s$-process yields in the interval from La to Gd, simultaneously solving the rate equations for 88 nuclides in this interval. The $n$-capture cross sections were adopted from Bao et~al.~(\cite{bao2000}) with some modifications such as the new data from Best et~al.~(\cite{best2001}). The $\beta$-decay rates and the electron-capture rates were from Takahashi \& Yokoi~(\cite{takahashi1987}). In stationary-flow calculations we found the abundances of Eu relative to La, Ce, Pr, Nd, Sm and Gd to stay significantly below the observed values and vary rather little when the neutron density increased from $10^8$ to $10^{14}$~cm$^{-3}$, with some tendencies to peak around $10^{12}$; only Eu/Sm and Eu/Pr came then close to the observed values. For the time-dependent case we varied neutron exposures and temperatures, but again in this case we could not recover the high observed Eu/X ratios for X= La, Nd, Ce, or Gd with any combination of parameters tried. Although these results are preliminary, and a number of nuclear rates are still rather uncertain, we find it less probable that a high-neutron density $s$-process would be responsible for the characteristics of the r+s stars.

\paragraph{VIII. A relation to blue stragglers?} The fraction of r+s stars in Table~\ref{Tstars} that have $T_\mathrm{eff} > 6000$~ K is fairly high, as compared with the fraction of hot s stars and r-II stars.   This may reflect small-number statistics or selection effects -- it is certainly more difficult to detect an r-II star than a r+s star at the turn-off point (TOP) due to the fact that the former, at least in our sample, tend to have both lower metallicities and lower [Eu/Fe].  Furthermore, since r+s stars typically have $\mathrm{[C/Fe]} > 2.0$ (see Table~\ref{Tstars}), the cooler r+s might have been rejected from the sample of Paper\,{\sc ii} due to their strong CH lines.
The dominance of detections of r+s stars at the TOP might also point at some relation between the r+s stars and the blue stragglers in globular clusters and in the field.  These objects, lying above and blueward of the turn-off in the colour-magnitude diagram of clusters, are generally believed to be the result of close-binary evolution with mass transfer, of stellar collisions or of mixing of the stellar interior (for a review, see, e.g.\ Trimble~\cite{trimble1993}). Sneden et~al.~(\cite{sneden2003b}) recently analysed spectra of six field blue stragglers with $\mathrm{[Fe/H]} < -2$.  Three of them were spectroscopic binaries and were found to have enhanced carbon and Sr and Ba. This was interpreted as the result of mass transfer from companions in the AGB stage. One of these stars was remarkably rich in Pb and had in fact a high, though uncertain, Eu abundance -- that star is CS\,29497-030 and is included as an r+s star in our Table~\ref{Tstars}. The idea that sub-giant CH stars may be related to blue stragglers was already put forward by Luck \&\ Bond (\cite{luck1991}). However, in order for this scenario to apply to the r+s stars, and not only as a provider of the $s$-elements, an idea which is already included in the scenarios III-VI above, one has to invoke a production mechanism for the $r$-elements. One might hope that the alternative origin of blue stragglers as the result of stellar collisions might provide not only overall mixing of CNO and He (e.g.\ Sills et~al.~\cite{sills2005}) but possibly also heavy neutron fluxes at head-on collisions, leading to formation of Eu and other $r$-elements. It is, however, highly uncertain whether such a speculative mechanism, presumably acting in the centre of dense globular clusters, could contribute enough r+s stars in the field.          

\paragraph{IX. Origin from a common cloud of r+s stars:}  In view of the similarity in chemical composition of the r+s stars it is not inconceivable that they were formed out of gas in a common cloud, and even were accreted by the Galaxy at a later stage. To explore this hypothesis we have calculated space velocities for the stars. In the absence of accurate distances for but one star (CS~31062-012), we used photometric distances estimated by reading off the colour-magnitude diagram of the metal-poor globular cluster M92 (Ruelas-Mayorga \& S{\'a}nchez~\cite{ruelas2005}) at the appropriate reddening-corrected $B-V$. Alternatively, we have compared to theoretical isochrones of Kim et~al.~(\cite{kim2002}) with $\mathrm{[Fe/H]}=-2.5$ and an age of 12~Gyr in the $T_\mathrm{eff}$-$\log g$ diagram and derived the absolute magnitudes from that. For the four stars with well known kinematic data we find space velocities with a spread of approximately 100 km/s in the $U$, $V$ and $W$ components, and conclude from this that there is no suggestion of a common space motion.

\section{Conclusions}
\label{conclusions}

In this paper we have analysed a high-quality spectrum of the $r$- and $s$-element rich star HE~0338$-$3945, deriving abundances for 33 elements and upper limits for an additional 6 elements.  We found the abundances of this star to be very similar to those of HE~2148$-$1247, which is supported by an homogeneous analysis of the two spectra.  In fact the abundance patterns among known r+s stars in the literature seem to be quite similar considering the differences between analyses.  Based on our results for HE~0338$-$3945 and our survey of other r+s stars, we have discussed a range of scenarios for the formation of the r+s stars.   Some scenarios have considerable merit, but all have also drawbacks or problems. Scenario I is already dismissed. Scenarios II and III need the ISM to be highly inhomogeneous to produce such high $r$-process abundances as detected in these stars.  We note that the [Eu/Fe] ratios for the r+s stars are systematically higher than those of the r-II stars which might suggest that a process with two independent steps, one of which is also responsible for the r-II stars is not very probable.  However, this is uncertain due to the possibly significant contribution of Eu by the $s$-process, and the possible relative effects of convective mixing between the r+s stars, which are predominantly TOP stars, and the r-II stars, which are all giants (see also discussion below).
Scenario II is also improbable as it requires all r+s stars to have passed at least the RGB phase.  Scenario III presently seems unlikely in view of the frequency of r+s stars and their apparent chemical homogeneity. The discussed modification of scenario III to include supernova-triggered binary formation seems plausible, though it is presently unclear whether such events can provide the observed frequency of r+s stars. Scenario IV is improbable as it requires strict stellar orbit constraints to produce the chemical homogeneity of the r+s class.   Scenarios V and VI may be considered, but as they depend heavily on physical processes and parameters that are poorly understood at present, these scenarios at best have to be regarded as uncertain. The observed chemical uniformity is not an obvious consequence of them.  Scenario VII still needs a site which can produce the high neutron densities at suitable exposures; the scenario also has to be supported by renewed and more realistic $s$-process calculations. It is also not clear how well this scenario would explain the chemical homogeneity. A probable merit of it is, however, that the observed great but varying amounts of Pb may result naturally from the high neutron fluxes. The absence of observed radial velocity variations for several stars (see Preston \& Sneden~\cite{preston2001} and Aoki et~al.~\cite{aoki2002}) is a problem, although not very severe as yet, for all scenarios III-VIII. Scenario VIII seems hypothetical and it is also questionable whether it is efficient enough. The available astrometric data, though admittedly meager, does not support Scenario IX.

The majority of the r+s stars in Table~\ref{Tstars} are TOP stars, while all the r-II stars are giants. One should note the great difference between the TOP stars and the red giants as regards two important aspects:  First, the convective zone of the TOP stars is less than 5\% of the stellar mass, while it is typically 75\% of the mass for the giants. Thus, the dilution of enriched material transferred to the stellar surface of a TOP star is at least a factor of 10 less than for a giant. So, if the enhancements of $n$-capture elements are due to mass transfer from a companion, the amount transferred has to be at least one order of magnitude greater for the r-II stars than for most of the TOP stars. Second, the space volume surveyed in apparent-magnitude limited surveys is at least a factor of 200 greater for the giants than for the r+s stars. This suggests that the space density of r+s stars is considerably higher than that of r-II stars. In spite of their scarcity, it is interesting that no r-II stars have been found at the TOP, since a much smaller transfer of $r$-elements would raise the atmospheric abundances to considerable values.  More statistically controlled surveys are desirable.   

Future detailed studies, with accurate analysis of all known and suspected stars, are worthwhile. Spectra with high $S/N$ and high resolution, also in the the UV, are important to obtain better limits on Ag, Th and U and to get results for $r$-elements that have lines in the UV such as Os and Ir. Attempts to measure Eu isotopic ratios for these stars should be continued. The analysis of certain important elements, such as Eu and Ag, should be scrutinised for NLTE and 3D effects.  More detailed calculations of the $s$-process at high neutron densities should be carried out. Finally, all known r+s stars should be monitored in an attempt to disclose their possible binarity.

\begin{acknowledgements}
We thank Nikolai Piskunov for his invaluable help during the reductions, and Oleg Kochukhov for his kind help with merging the spectra.  We are grateful to Kjell Eriksson for computing MARCS models, as well as certain checks of the effective temperature.  Nils Ryde and Rurik Wahlin are thanked for providing molecular line lists.  We are grateful to Judith Cohen for providing us with her spectrum of HE~2148$-$1247.  Roberto Gallino and Sara Bisterzo are thanked for valuable discussions.  We made use of the NASA ADS and VALD databases. 

The Uppsala group acknowledges the support of the Swedish Research Council (VR).  NC acknowledges the support of Deutsche Forschungsgemeinschaft (grant Ch~214/3-1).  TCB acknowledges partial support from grants AST 04-06784, and PHY 02-16783, Physics Frontier Centers/JINA: Joint Institute for Nuclear Astrophysics, awarded by the US National Science Foundation. 

\end{acknowledgements}


\appendix

\section{Line List} 
\label{appendix}

The complete line list is presented in Table 2, which is available only electronically.  Here we briefly comment on line selection issues and data sources for each element, particularly for the oscillator strengths, hfs and isotopic splitting.  Other data such as wavelengths, excitation potentials were collected from various sources, often the Vienna Atomic Line Database (VALD, Kupka et~al.~\cite{kupka1999}), the NIST compilations or Sneden et~al.~(\cite{sneden2003}).

\paragraph{Carbon:} The abundance of carbon was derived from the A--X bands at 4310--4313\,{\AA} (0,0) and 4362--4367\,\AA (1,1), along with two B$^2\Sigma-$X$^2\Pi$ (1,0) CH lines at 3638\,{\AA} and 3661\,{\AA}. The band regions used in this programme are the same as those used by Paper\,{\sc ii}.  The two single lines were selected from the list of J{\o}rgensen et~al.~(\cite{jorgensen1996}) as they are unblended, of moderate strength, and have reasonable $S/N$ in the observed spectrum.  The C abundance was derived from $^{12}$CH features assuming an isotopic ratio of $^{12}\mathrm{C}/^{13}\mathrm{C}=\infty$ as in Paper\,{\sc ii}.

\paragraph{Nitrogen:} The nitrogen abundance was derived using the B$\Sigma-$X$\Sigma$ bands of CN at 3872\,{\AA} (1,1) and 3882\,{\AA} (0,0). The line data was extracted from a list made available by Plez (1998, private communication) and is described by Hill et~al.~(\cite{hill2002}). 

\paragraph{Oxygen:} The data for the oxygen lines were compiled from VALD, except for the $\log gf$ values which were astrophysical values taken from Jonsell et~al.~(\cite{jonsell2005}). The differences from the NIST $\log gf$ values in VALD were less than 0.02\,dex.

\paragraph{Sodium:} For two of the three \ion{Na}{i} lines the $f$-values of Wiese \& Martin~(\cite{wiese1980}) and hfs from McWilliam et~al.~(\cite{mcwilliam1995}) were used.  The $f$-value for the 8183\,{\AA} line was taken from VALD. No hfs data were available for this line, and as it is weak this should not be important.

\paragraph{Magnesium:} A total of 7 lines of \ion{Mg}{i} were used.  Data for the lines at 3829, 4571, and 4703\,{\AA} were taken from Paper\,{\sc ii}.  Due to contamination by CH, Balmer lines, and some residual reduction artefacts not all lines from Paper\,{\sc ii} could be used.  Oscillator strengths for the lines at 5173, 5184, and 5528\,{\AA} originate from Wiese \& Martin~(\cite{wiese1980}).  For the 5711 \,{\AA} line the astrophysical $\log gf$-value from Jonsell et~al.~(\cite{jonsell2005}) was adopted.

\paragraph{Aluminium:} The \ion{Al}{i} resonance line at 3961\,{\AA} was used for the analysis. The hfs was taken from McWilliam et~al.~(\cite{mcwilliam1995}). 

\paragraph{Calcium:} The $f$-values for the 10 \ion{Ca}{i} lines were taken from Wiese \& Martin~(\cite{wiese1980}), Smith~(\cite{smith1981}), Jonsell et~al.~(\cite{jonsell2005}), and VALD.

\paragraph{Scandium:} The abundance of scandium was determined from 5 \ion{Sc}{ii} lines, using $f$-values from Lawler \& Dakin~(\cite{lawler1989}). Hfs for the lines 4247 and 4415\,{\AA} was taken from McWilliam et~al.~(\cite{mcwilliam1995}).  The remaining lines are very weak and hfs was not considered.

\paragraph{Titanium:} A total of 11 lines of \ion{Ti}{i} and 13 of \ion{Ti}{ii} were analysed. The oscillator strengths were gathered from various sources: Blackwell et~al.~(\cite{blackwell1982a}), Blackwell et~al.~(\cite{blackwell1982b}), Fuhr \& Wiese~(\cite{fuhr1996}), Grevesse et~al.~(\cite{grevesse1989}), Martin et~al.~(\cite{martin1988}), Pickering et~al.~(\cite{pickering2001}), Ryabchikova et~al.~(\cite{ryabchikova1994}) and VALD. 

\paragraph{Vanadium:} The vanadium abundance is based on one \ion{V}{i} and 4 \ion{V}{ii} lines. The $f$-value data were taken from Fuhr \& Wiese~(\cite{fuhr1996}).

\paragraph{Chromium:} The $f$-values for the 6 \ion{Cr}{i} and 2 \ion{Cr}{ii} lines were taken from Fuhr \& Wiese~(\cite{fuhr1996}) and Martin et~al.~(\cite{martin1988}). 

\paragraph{Manganese:} For the two lines of \ion{Mn}{i} we adopted the $\log gf$ data from Booth et~al.~(\cite{booth1984}), and hfs was computed using the data compiled in Lef\`ebvre et~al.~(\cite{lefebvre2003}).  For the three \ion{Mn}{ii} lines we used data from Kling \& Griesmann~(\cite{kling2000}), with hfs from Holt et~al.~(\cite{holt1999}). 

\paragraph{Iron:} A total of 61 \ion{Fe}{i} lines and 8 \ion{Fe}{ii} lines were analysed. The neutral line oscillator strengths are taken from O'Brian et~al.~(\cite{obrian1991}), Fuhr \& Wiese~(\cite{fuhr1996}), Bard et~al.~(\cite{bard1991}), Bridges \& Kornblith~(\cite{bridges1974}), and VALD.  For the \ion{Fe}{ii} lines $f$-values were adopted from Bi\'emont et~al.~(\cite{biemont1991}), Fuhr \& Wiese~(\cite{fuhr1996}), Heise \& Kock~(\cite{heise1990}), Kroll \& Kock~(\cite{kroll1987}), Moity~(\cite{moity1983}), Schnabel et~al.~(\cite{schnabel2004}), and VALD. 

\paragraph{Cobalt:} The cobalt abundance was derived from 4 \ion{Co}{i} lines using $f$-values from Cardon et~al.~(\cite{cardon1982}) and Nitz et~al.~(\cite{nitz1999}), and hfs data from Pickering~(\cite{pickering1996}).

\paragraph{Nickel:} The 8 lines of \ion{Ni}{i} have oscillator strengths from Blackwell et~al.~(\cite{blackwell1989}), Fuhr et~al.~(\cite{fuhr1988}), Fuhr \& Wiese~(\cite{fuhr1996}), Huber \& Sandeman~(\cite{huber1980}), and VALD.

\paragraph{Copper:} We analysed two lines of \ion{Cu}{i}, both in the UV. The $\log gf$ data were taken from Fuhr \& Wiese~(\cite{fuhr1996}), and hfs data from Kurucz~(\cite{kurucz1995}).

\paragraph{Strontium:} One \ion{Sr}{i} line and 4 \ion{Sr}{ii} lines were analysed. Hyperfine and isotopic splitting are accounted for in the 4077 and 4215~\AA\ lines (see discussion in Paper\,{\sc ii}).  We did not find hyperfine structure data for the remaining lines; however, as they are weak, this should not be important. The $f$-values were gathered from Migdalek \& Baylis~(\cite{migdalek1987}), Pinnington et~al.~(\cite{pinnington1995}), and Sneden et~al.~(\cite{sneden2003}). 

\paragraph{Yttrium:} The $f$-values for the 11 lines of \ion{Y}{ii} are from Hannaford et~al.~(\cite{hannaford1982}). 

\paragraph{Zirconium:} All but two of the 16 lines of \ion{Zr}{ii} were analysed with $f$-values from Bi\'emont et~al.~(\cite{biemont1981}). The remaining lines at 3356 and 4019\,{\AA} have $f$-values from VALD. 

\paragraph{Silver:} The upper limit to the silver abundance was derived from two \ion{Ag}{i} lines. The oscillator strengths were taken from from Fuhr \& Wiese~(\cite{fuhr1996}).  The hfs data was taken from Ross \& Aller~(\cite{ross1972}). 

\paragraph{Barium:} The barium abundance is derived using three \ion{Ba}{ii} lines. The $\log gf$ data came from Gallagher (\cite{gallagher1967}) and VALD.  The isotopic splitting and hfs were taken from Villemoes et~al.~(\cite{villemoes1993}) for the two weak lines, noting that data for only the lower levels are available.  For the 4554\,{\AA} line, hfs and isotopic splitting from McWilliam~(\cite{mcwilliam1998}) was adopted.  In all cases, the solar isotopic composition from Anders \& Grevesse~(\cite{anders1989}) was assumed. We note that if a pure $r$-process isotopic composition is assumed we find an abundance only 0.03~dex lower.

\paragraph{Lanthanum:} The 8 lines of \ion{La}{ii} were analysed employing oscillator strengths and hfs data from Lawler et~al.~(\cite{lawler2001a}). For the lines at 4921 and 4922\,{\AA} only the hfs of the lower levels was considered, as no data were available for the upper levels. 

\paragraph{Cerium:} Abundances were derived from 15 lines of \ion{Ce}{ii}. The $f$-values were gathered from various sources, Palmeri et~al.~(\cite{palmeri2000}), Sneden et~al.~(\cite{sneden1996}), Gratton \& Sneden~(\cite{gratton1994}), and VALD.  

\paragraph{Praseodymium:} The $\log gf$ values for the 4 lines of \ion{Pr}{ii} were taken from Ivarsson et~al.~(\cite{ivarsson2001}).

\paragraph{Neodymium:} A total of 34 \ion{Nd}{ii} lines were analysed. The oscillator strengths were taken from Den Hartog et~al.~(\cite{denhartog2003}). 

\paragraph{Samarium:} The 4 lines of \ion{Sm}{ii} were analysed using $f$-values from Bi\'emont et~al.~(\cite{biemont1989}).

\paragraph{Europium:} The abundance of europium is based on 3 \ion{Eu}{ii} lines, with $f$-values, hfs and isotopic splitting taken from Lawler et~al.~(\cite{lawler2001c}). Solar $r$-process isotopic fractions were assumed.

\paragraph{Gadolinium:} The $f$-values for the 8 \ion{Gd}{ii} lines were collected from Sneden et~al.~(\cite{sneden2003}); the paper of Cowan et~al.~(\cite{cowan2002}) should be consulted for details.

\paragraph{Terbium:} The $f$-value for the single line of \ion{Tb}{ii} was taken from Lawler et~al.~(\cite{lawler2001d}).

\paragraph{Dysprosium:} The $f$-values for 15 of the 16 \ion{Dy}{ii} lines analysed were taken from Wickliffe et~al.~(\cite{wickliffe2000}). For the line at 3944\,{\AA} the $f$-value is from from Bi\'emont \& Lowe~(\cite{biemont1993}).

\paragraph{Holmium:} The single \ion{Ho}{ii} line at 4152\,{\AA} has atomic data including hfs taken from Sneden et~al.~(\cite{sneden2003}).

\paragraph{Erbium:} The $f$-values for the 3 \ion{Er}{ii} lines were taken from Musiol \& Labuz~(\cite{musiol1983}) and Kurucz~(\cite{kurucz1995}).

\paragraph{Thulium:} The 4 \ion{Tm}{ii} lines have oscillator strengths from Kurucz~(\cite{kurucz1995}). According to Sneden et~al.~(\cite{sneden1996}) these are rescaled laboratory data from Corliss \& Bozman~(\cite{corliss1962}).

\paragraph{Ytterbium:} The $f$-values used for our two \ion{Yb}{ii} lines are from Sneden et~al.~(\cite{sneden2003}). They have renormalised theoretical oscillator strengths from Bi\'emont et~al.~(\cite{biemont1998}) using lifetimes from Pinnington et~al.~(\cite{pinnington1997}). The isotopic and hyperfine splitting data are taken from M{\aa}rtensson-Pendril et al.~(\cite{martensson1994}), and the solar isotopic ratios from Anders \& Grevesse~(\cite{anders1989}) have been assumed.

\paragraph{Lutetium:} For the single \ion{Lu}{ii} line we adopted the $\log gf$ and hfs data from Sneden et~al.~(\cite{sneden2003}).  The oscillator strength is from Quinet et~al.~(\cite{quinet1999}), renormalised using lifetimes from Fedchak et~al.~(\cite{fedchak2000}).

\paragraph{Hafnium:} We have one line of \ion{Hf}{ii} with the $\log gf$ taken from VALD.

\paragraph{Lead:} The abundance is derived from two \ion{Pb}{i} lines.  The oscillator strengths were taken from Bi\'emont et~al.~(\cite{biemont2000}) and the hfs and isotopic splitting data from Manning et~al.~(\cite{manning1950}).  Solar isotopic ratios from Anders \& Grevesse~(\cite{anders1989}) have been assumed.

\paragraph{Thorium:} An upper limit for the thorium abundance is derived from the \ion{Th}{ii} line at 4019.129\,{\AA} employing the oscillator strength from Nilsson et~al.~(\cite{nilsson2002b}). 

\paragraph{Uranium:} The $f$-value for the uranium line is taken from Nilsson et~al.~(\cite{nilsson2002a}).

\end{document}